\documentclass[aps,prl,twocolumn,amsmath,amssymb,showpacs,superscriptaddress,reprint]{revtex4-2}
\pdfoutput=1
\usepackage[T1]{fontenc}
\usepackage{graphicx}
\makeatletter

\usepackage[english]{babel}
\usepackage[T1]{fontenc}
\usepackage[utf8]{inputenc}
\IfFileExists{lmodern.sty}{\usepackage{lmodern}}{}
\usepackage{color}
\definecolor{darkblue}{rgb}{0,0,0.6}
\definecolor{darkred}{rgb}{0.6,0,0}
\definecolor{darkgreen}{rgb}{0,0.6,0}
\usepackage{setspace}
\usepackage{environ}
\usepackage{wasysym}
\usepackage{times}
\usepackage[colorlinks=true,urlcolor=darkblue,citecolor=darkblue,linkcolor=darkred,hyperindex=true,hyperfootnotes=false]{hyperref}
\makeatletter
\def\maketitle{
\@author@finish
\title@column\titleblock@produce
\suppressfloats[t]}
\makeatother

\newcommand{\CF}{C_\mathcal{F}}

\begin{document}
\title{The Anomalous Transport of Tracers in Active Baths}

\author{Omer Granek}
\affiliation{Department of Physics,
Technion-Israel Institute of Technology,
Haifa, 3200003, Israel.}
\author{Yariv Kafri}
\affiliation{Department of Physics,
Technion-Israel Institute of Technology,
Haifa, 3200003, Israel.}
\author{Julien Tailleur}
\affiliation{Universit{\' e} de Paris, Laboratoire Matière et Systèmes Complexes (MSC),
UMR 7057 CNRS, F-75205 Paris, France.
}

\begin{abstract}
We derive the long-time dynamics of a tracer immersed in a one-dimensional active bath. In contrast to previous studies, we find that the damping and noise correlations possess long-time tails with exponents that depend on the tracer symmetry. For generic tracers, shape asymmetry induces ratchet effects that alter fluctuations and lead to superdiffusion and friction that grows with time when the tracer is dragged at a constant speed. In the singular limit of a completely symmetric tracer, we recover normal diffusion and finite friction. Furthermore, for small symmetric tracers, the active contribution to the friction becomes negative: active particles enhance motion rather than oppose it. 
These results show that, in low-dimensional systems, the motion of a passive tracer in an active bath cannot be modeled as a persistent random walker with a finite correlation time.
\end{abstract}

\maketitle
Since Einstein and Smoluchowski, the motion of a tracer particle in a bath has been a topic of much interest~\cite{Hanggi2005}. The simplest textbook framework   models the motion of the particle as a memoryless Brownian motion using an underdamped Langevin equation~\cite{Gardiner1985,Chaikin2000,Kardar2007}. The momentum autocorrelation function then decays exponentially with a single timescale, signaling a transition between inertial and viscous regimes. This was, however, found to be oversimplistic: the conservation of momentum in the solvent instead leads to a power-law decay~\cite{Alder1967,Pomeau1975,VanBeijeren1982} and a host of interesting phenomena---especially in low dimensions---such as the breakdown of the Fourier law~\cite{Kirkpatrick2002,Dhar2008,Spohn2016}.

When compared with the equilibrium case, active fluids reveal a much richer physics, from the ratchet effects induced by asymmetric gears~\cite{DiLeonardo2010,Sokolov2010,Kaiser2014,Maggi2015} and rectifiers~\cite{Galajda2007,Wan2008,Tailleur2009,Stenhammar2016,Reichhardt2017,Speck2020} to the long-ranged forces and currents generated by asymmetric obstacles~\cite{Nikola2016,Baek2018,Granek2020,Speck2020,Speck2021}. Over the past two decades, much activity has been devoted to studying passive tracers in active baths~\cite{Wu2000,Maggi2014,Argun2016,Maggi2017a,Chaki2018,Chaki2019,Dabelow2019,Goswami2019,Knezevic2020,Ye2020,Belan2021,Soni2003,Chen2007,Loi2008,Underhill2008,Lau2009,Leptos2009,Dunkel2010,Kurtuldu2011,Mino2011,Zaid2011,Foffano2012,Mino2013,Kasyap2014,Morozov2014,Thiffeault2015,Patteson2016,Suma2016,Burkholder2017,Jerez2017,Kurihara2017,Pietzonka2018,Burkholder2019a,Chatterjee2019,Dabelow2019,Goswami2019,Kanazawa2020,Ye2020,Knezevic2020,Reichert2020,Abbaspour2021,Belan2021,Katuri2021,Reichert2021}.
In the adiabatic limit in which the bath's relaxation is much faster than
the tracer's response~\cite{Green1952,Mori1965,Kubo1966,VanKampen1985,VanKampen1986,SeifertRPP2012,Maes2015,Kruger2017}, the tracer's dynamics is described by a generalized Langevin equation. In 1D, it reads as
\begin{align}
\gamma_{0}\dot{X}(t)+\int_{0}^{t}dt'\gamma (t-t')\dot{X}(t') & =\mathcal{F}(t)+\eta(t)~,\label{eq:GLE}
\end{align}
where the interactions with the active particles lead to a stochastic force $\mathcal{F}(t)$ and a retarded friction $\int_{0}^{t}dt'\gamma (t-t')\dot{X}(t')$. Equation~\eqref{eq:GLE} also includes a memoryless viscous medium at temperature $T$ that leads to the friction coefficient $\gamma_0$ and a Gaussian white noise $\eta(t)$ satisfying $\left\langle \eta(t)\eta(t') \right\rangle = 2\gamma_0 T \delta(t-t')$.  Despite many efforts, a single unifying picture for the friction $\gamma(t)$ and the force-force correlation functions $C_{\mathcal{F}}\equiv \langle \mathcal{F}(t) \mathcal{F}(0) \rangle_\text{c}$ does not emerge from the existing results. 

First, a large class of experimental and numerical studies
has suggested that the random, finite-duration encounters between the bath particles and the tracer lead to an exponential decay of $\gamma(t)$ over a short timescale~\cite{Wu2000,Maggi2014,Argun2016,Maggi2017a,Chaki2018,Chaki2019,Dabelow2019,Goswami2019,Knezevic2020,Ye2020,Belan2021}. Equation~\eqref{eq:GLE} then reduces to $(\gamma_0+\gamma_\text{T}) \dot{X}(t)=\mathcal{F}(t)$, where $\gamma_\text{T}\equiv\int_{0}^{\infty}dt\gamma (t)$. 
In this case, similarly to an underdamped Brownian particle, the large-scale motion of the tracer is  diffusive. This has been justified analytically in the simple case of a tracer connected by linear springs to a bath of active Ornstein-Uhlenbeck particles~\cite{Maes2020}---an active counterpart to the celebrated work of Vernon and Feynman~\cite{Feynman1963,Ford1965,Caldeira1983}.

In contrast, a second class of experiments and models on so-called wet-active matter suggests a more complex physics~\cite{Chen2007,Leptos2009,Kurtuldu2011,Zaid2011,Thiffeault2015,Kurihara2017,Kanazawa2020}. The long-ranged decay of hydrodynamic interactions can indeed turn $\gamma(t)$ and $C_{\mathcal{F}}(t)$ into power laws~\cite{Chen2007,Zaid2011,Kanazawa2020}. These may lead to anomalous diffusion on intermediate timescales but, ultimately, lead to long-time diffusion.

We note, however, that long-time tails are generic, even in the absence of hydrodynamic interactions. Indeed, the fluctuating density of active particles is a conserved quantity---and hence a slow field---so that the bath cannot have a single characteristic relaxation time. This leads to power-law memory
and correlations, as already noted for equilibrium~\cite{Pomeau1975,Hanna1981,VanBeijeren1982,Boon1991} and nonequilibrium~\cite{Goychuk2002,Lau2003} systems, including phoretic colloids~\cite{Golestanian2009} and driven tracers~\cite{Benichou2013,Illien2013}.
In low-dimensional systems, these tails may result in anomalous transport over long timescales~\cite{Golestanian2009,Benichou2013}. Although thoroughly studied in other contexts, these effects were so far  overlooked for tracers in dry active baths.

\begin{figure}
\includegraphics[width=1\columnwidth]{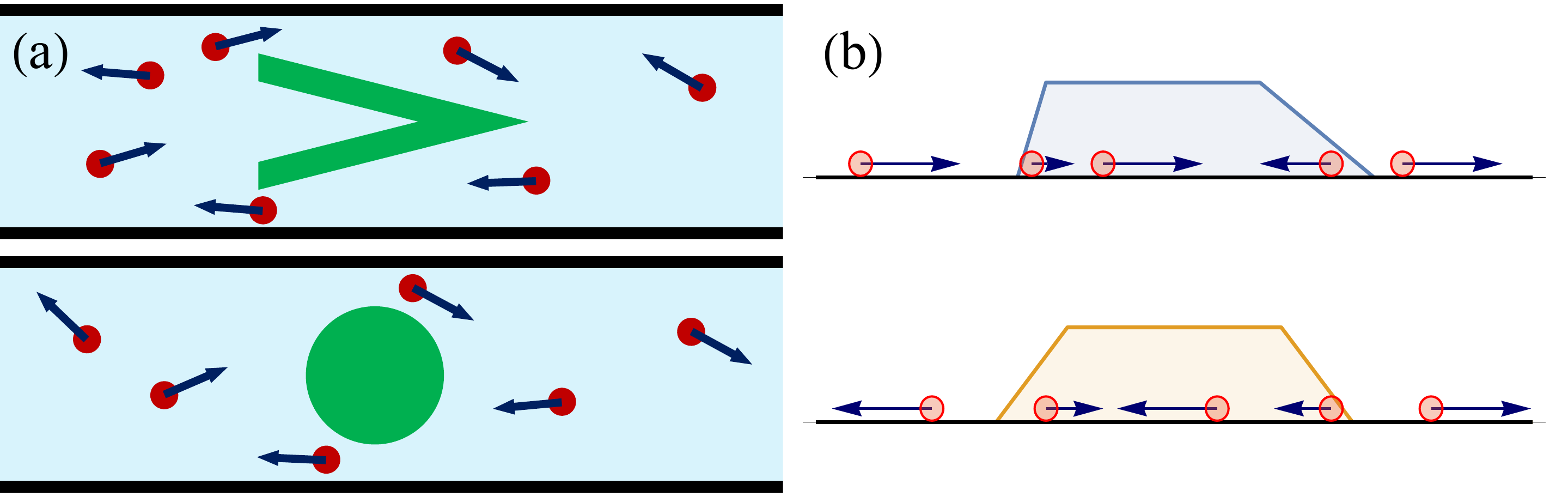}
\caption{(a) A large tracer and a bath of small active particles are immersed in a viscous medium inside a long narrow channel. (b) The short transverse dimension allows one to model the channel as a one-dimensional system where particles can bypass each other and the tracer, even though the transverse and orientational fluctuations of the tracer are lost in this one-dimensional description. Top: asymmetric tracer. Bottom: symmetric tracer. \label{fig:setup}}
\end{figure}

In this Letter, to resolve this issue, we consider the simplest nontrivial system in which Eq.~\eqref{eq:GLE} can be systematically derived: a single tracer immersed in a dry one-dimensional active bath of run-and-tumble particles. To remain as close as possible to the phenomenology of an active bath in $d>1$ dimensions, we allow particles to overtake each other and the tracer, hence modeling the latter by a soft repulsive potential $V(x)$, see Fig.~\ref{fig:setup}. 
Starting from the coupled dynamics of the bath particles and tracer positions, $\{x_{i}(t),X(t)\}$, we determine explicitly the long-time behaviors of $\gamma(t)$ and $C_\mathcal{F}(t)$ as functions of the tracer shape and of the microscopic parameters of our model. To do so, we employ a \emph{controlled} adiabatic expansion~\cite{DAlessioJSTAT2016,Weinberg2017} valid in the large $\gamma_0$ limit in which the tracer dynamics can be described by Eq.~\eqref{eq:GLE}. Our results show the emergence of long-time tails that lead to interesting and qualitatively different behaviors for symmetric and asymmetric tracers. 
For generic, \emph{asymmetric} tracers, ratchet effects make $\gamma(t)$ and $C_\mathcal{F}(t)$ scale as $\sim t^{-1/2}$ in the long-time limit, leading to \emph{superdiffusive} behavior around their mean displacements:
\begin{align}
\left\langle X^{2}(t)\right\rangle _{\text{c}} \equiv \langle X^2(t) \rangle -\langle X(t) \rangle^2 & \sim Kt^{3/2}~.\label{eq:supdiff}
\end{align}
When the tracer is towed at a constant velocity $U$, it experiences a friction force from the active particles that grows as:
\begin{align}
\frac{f_{\text{fric}}(t)}{U} & \sim-\Gamma_{\text{T}}t^{1/2}~.
\end{align}
We provide below explicit expressions for $K$ and $\Gamma_\text{T}$ in the presence of a soft asymmetric potential in a dilute active bath. 
In the singular limit of a \emph{symmetric} tracer, $C_\mathcal{F}(t)$ and $\gamma(t)$ scale as $\sim t^{-3/2}$, similar to a tracer in a bath of equilibrium Brownian particles~\cite{Hanna1981,Hanna1982}, which yields a \emph{diffusive} behavior:
\begin{align}
\left\langle X^{2}(t)\right\rangle _{\text{c}} & \sim2Dt~.\label{eq:diff}
\end{align}
Towing the tracer at constant velocity $U$, the active particles exert a \emph{finite} friction force:
\begin{align}
\frac{f_{\text{fric}}(t)}{U} & =-\gamma_{\text{T}}-\gamma_{1}t^{-1/2}+\mathcal{O}(t^{-3/2})~,\label{eq:fric}
\end{align}
where $\gamma_\text{T}\equiv\int_0^\infty dt \gamma(t)$. Interestingly, for small tracer sizes, $\gamma_{\text{T}}$ and $\gamma_1$ are negative: the active bath pushes the tracer in the towing direction. We provide perturbative expressions for $D$ and $\gamma_\text{T}$ and defer their systematic derivations for later work~\cite{OmerFuture}. 
All our results are confirmed by microscopic simulations shown in Fig.~\ref{fig:Simulation-results-for}. The derivation presented below suggests that the exponents are \emph{universal} to any bath with long-time diffusive statistics. We confirm that they hold in the presence of soft repulsive interparticle forces in Sec. I of the Supplemental Material~\cite{SM}.
\begin{figure}
\includegraphics[width=1\columnwidth]{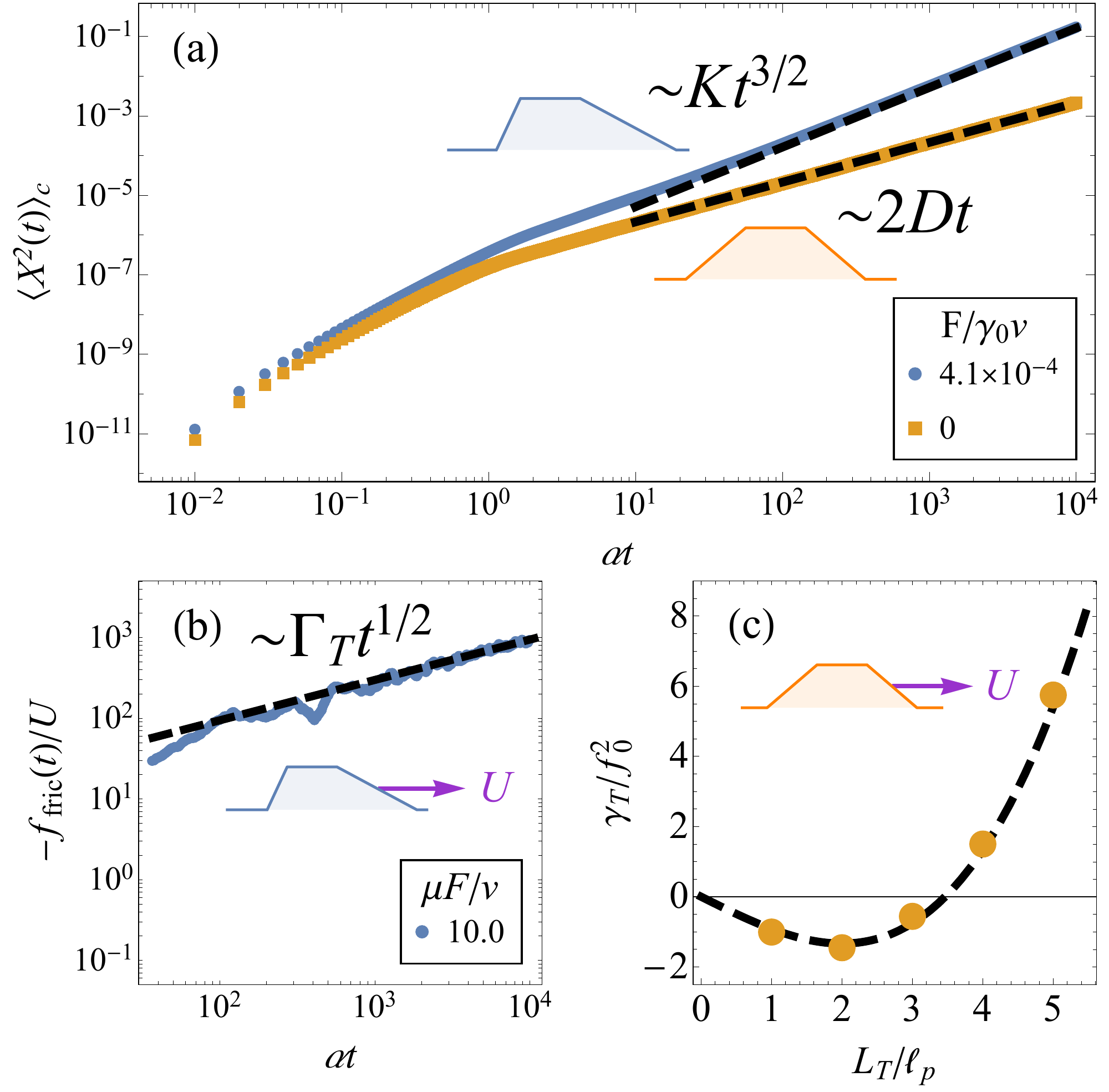}
\caption{Simulation results (symbols) compared with our
theoretical predictions for the long-time limit, without any fitting parameters, (dashed black lines): (a) mean squared displacement for symmetric and asymmetric tracers; (b) friction force exerted on an asymmetric tracer; (c) symmetric-tracer friction coefficient vs tracer size $L_\text{T}$. Simulation details and results for soft repulsive interactions are given in Sec. I of the Supplemental Material~\cite{SM}. \label{fig:Simulation-results-for}}
\end{figure}

\emph{Model.}—We consider bath particles moving with speed $v$ and randomly switching their orientations with rate $\alpha/2$, leading to a persistence length $\ell_{\text{p}}=v/\alpha$. The tracer interacts with the active bath via a short-range potential $V$ which vanishes outside $[0,L_\text{T}]$, such that the force on bath particle $i$ is $f(x_{i}-X)=-\partial_{x_{i}}V(x_{i}-X)$ and the tracer size is $L_\text{T}$. We take $\left|\mu f(x)\right|<v$ so that particles are able to cross the tracer, which emulates the channel in Fig.~\ref{fig:setup}(a). The tracer and bath-particle dynamics thus read as
\begin{align}
\gamma_{0}\dot{X}(t) & =F_\text{tot}(t)\equiv-\sum_{i}f[x_{i}(t)-X(t)]~,\label{eq:body}\\
\dot{x}_{i}(t) & =v\sigma_i (t)+\mu f[x_{i}(t)-X(t)]~,\label{eq:sde}
\end{align}
where the $\sigma_i (t)\in\{\pm1\}$ flip independently with rate $\alpha/2$ and $\mu$ is the bath-particle mobility. In Eqs.~\eqref{eq:body} and~\eqref{eq:sde} we neglected the thermal noises acting on the tracer and bath particles, which are typically much weaker than the active and viscous forces~\cite{Wu2000,Loi2008,Cates2012,Maggi2014,Ye2020}. (see Sec. V of the Supplemental Material~\cite{SM} for a discussion of the $T\neq 0$ case.) In the analytical derivations below we consider a dilute bath of active particles, without interparticle forces, in either infinite systems or periodic ones of size $L\gg L_\text{T},\ell_\text{p}$.


\emph{Theory.}—The fluctuating force $F_\text{tot}(t)$ differs from the average force $F$ exerted on a tracer held \emph{fixed}. This is due to both the tracer's motion and the stochasticity of the active bath. The average correction due to the tracer motion is characterized by $\gamma(t)$ in Eq.~\eqref{eq:GLE}. Within an adiabatic perturbation theory $\gamma(t)$ is defined as
\begin{align}
\left\langle F_\text{tot}(t)\right\rangle-F & \equiv -\int_0^t dt' \gamma(t-t')\dot{X}(t')~,\label{eq:df}
\end{align}
where the average is conditioned on a given realization of $\dot{X}(t)$. The fluctuations of $F_\text{tot}$ are then characterized through
\begin{align}
    \mathcal{F}(t) \equiv F_\text{tot}(t)+\int_0^t dt' \gamma(t-t')\dot{X}(t')\;.
\end{align}
Adiabatic perturbation theory tells us that, when $\gamma_0$ is large, the statistics of $\mathcal{F}(t)$ are identical to those of the force exerted on a tracer held fixed~\cite{Weinberg2017}. Furthermore, it relates $\gamma(t)$ and $\mathcal{F}(t)$ through an Agarwal-Kubo-type formula~\cite{DAlessioJSTAT2016}
\begin{align}
    \gamma (t-t') & =\left\langle \mathcal{F}(t)\partial_{X_0}\ln\rho_{\text{s}}\big[x(t')-X_0,\sigma(t')\big]\right\rangle^\text{s}~.\label{eq:Kgen}
\end{align}
Here, $\rho_{\text{s}}(x-X_0,\sigma)$ is the steady-state density of bath particles with orientation $\sigma$ and displacement $x-X_0$ from a tracer held fixed at $X_0$. The brackets $\langle\cdot\rangle^\text{s}$ represent an average with respect to $\rho_\text{s}$. In the following, we set $X_0=0$ without loss of generality. For an equilibrium bath at temperature $T$, $\left\langle \mathcal{F}(t)\right\rangle^\text{s}=0$
and Eq.~\eqref{eq:Kgen} reduces to the fluctuation-dissipation theorem (FDT) $\gamma(t)=C_\mathcal{F}(t)/T$ where $C_\mathcal{F}(t)=\langle\mathcal{F}(t)\mathcal{F}(0)\rangle^\text{s}$. Outside equilibrium, these constraints need not hold.

To characterize the tracer dynamics, we compute independently $F=\left\langle \mathcal{F}(t)\right\rangle^\text{s}$,
$C_\mathcal{F}(t)$ and $\gamma (t-t')$. To do so, we start from the expression for the steady state of noninteracting run-and-tumble particles in the presence of an external force $f(x)$~\cite{Kitahara1979,Solon2015PressureFluids}:
\begin{align}
\rho_{\text{s}}(x,\sigma)\! & =\!\frac{\frac{1}{2}\rho_{\text{L}}}{1\!+\!\sigma\frac{\mu}{v}f(x)}\!\exp{\!\left\{\beta_{\text{eff}}\!\int^{x}_0\!dy\frac{f(y)}{1\!-\!\left[\frac{\mu}{v}f(y)\right]^{2}}\right\}}\!,\label{eq:st}
\end{align}
where $\rho_{\text{L}}$ is the particle density at $x=0^{-}$, $T_\text{eff}=v^{2}/\mu\alpha$
is the effective temperature, and $\beta_{\text{eff}}=1/T_{\text{eff}}$. The steady-state density
is $\rho_{\text{s}}(x)=\sum_{\sigma}\rho_{\text{s}}(x,\sigma)$.

\emph{Asymmetric tracer.}—For an asymmetric tracer, the densities of active particles $\rho_\text{R}$ and $\rho_\text{L}$ at the right and left ends of the tracer differ and are given by $\rho_{\text{R}}=2\rho_0/\left[1+\exp{(\beta_\text{eff} \varepsilon)}\right]$ and  $\rho_{\text{L}}=2\rho_0/\left[1+\exp{(-\beta_\text{eff} \varepsilon)}\right]$, where $\varepsilon\equiv-\int dxf(x)/\{1-\left[\mu f(x)/v\right]^{2}\}$. The density difference then leads to a nonvanishing average force $F=-\int dxf(x)\rho_{\text{s}}(x)$ exerted on the tracer~\cite{Angelani2010,Mallory2014,Nikola2016}, which is given by
\begin{align}
F & =-T_\text{eff}(\rho_{\text{R}}-\rho_{\text{L}})=2T_{\text{eff}}\rho_{0}\tanh\left(\frac{\varepsilon}{2T_\text{eff}}\right)~,\label{eq:pump}
\end{align}
where we have introduced the average background
density $\rho_{0}=(\rho_{\text{R}}+\rho_{\text{L}})/2$. Note that Eq.~\eqref{eq:pump} is consistent with the ideal gas law applied to the left and right sides of the tracer.

The long-time behavior of $C_{\mathcal{F}}(t)$ and  $\gamma(t)$ can be derived from the knowledge of the propagator
$p(x,\sigma,t|x',\sigma',0)$. In the long-time limit, the dynamics of the active particles are diffusive so that the support of $p(x,\sigma,t|x',\sigma',0)$ spreads over a region of length $2b(t)$ around $x'$, where $b(t)\!\sim\!(\pi D_\text{eff} t)^{1/2}$ is a diffusive propagating front. For any $x\!-\!x'\!\ll\! b(t)$, and to leading order in $b(t)$, $p(x,\sigma,t|x',\sigma',0)$
has relaxed to the normalized steady-state distribution $\rho_{\text{s}}(x,\sigma)/\sum_{\sigma}\int_{-b(t)}^{b(t)}dx\rho_{\text{s}}(x,\sigma)$. For $L_\text{T}\ll2b(t)$, one can neglect the region inside the tracer
in the integral so that $\sum_{\sigma}\int_{-b(t)}^{b(t)}dx\rho_{\text{s}}(x,\sigma)\sim(\rho_{\text{R}}+\rho_{\text{L}})b(t)$, up to corrections of order $\mathcal{O}(L^{-1})$.
Since $b(t)\sim(\pi D_{\rm eff}t)^{1/2}$ we get
\begin{align}
p(x,\sigma,t|x',\sigma',0) & \sim\frac{\rho_{\text{s}}(x,\sigma)}{\rho_{\text{R}}+\rho_{\text{L}}}(\pi D_\text{eff}t)^{-1/2}~.\label{eq:propg}
\end{align}
This heuristic result can be derived exactly, within the adiabatic limit, and its subleading correction can be shown to scale as  $\mathcal{O}(t^{-3/2})$ (See Sec. II of the Supplemental Material~\cite{SM}).

On long times, $p(x,\sigma,t|x',\sigma',0)$ is
independent of the initial coordinate $(x',\sigma')$. Therefore,
two-point correlations are factorized in this limit. Furthermore, for $N$ noninteracting particles, the forces exerted by different particles on the tracer are uncorrelated so that $\CF(t)=N\{\left\langle f(t)f(0)\right\rangle^\text{s}-[\left\langle f(t)\right\rangle^\text{s}]^2\}$,  where $f(t)$ is the force due to a single bath particle. Since $\left\langle f(t)\right\rangle^\text{s}=F/N$, $N [\left\langle f(t)\right\rangle^\text{s}]^2$ only contributes a correction of order $\mathcal{O}(L^{-1})$ to $\CF(t)$. Using Eq.~\eqref{eq:propg}, $\CF(t)$ can then be evaluated as:
\begin{align}
\CF(t)\!& =\!\sum_{\sigma\sigma'}\!\int \!dxdx'\!f(x)p(x,\!\sigma\!,\!t|x'\!,\!\sigma'\!,\!0)f(x')\rho_{\text{s}}(x'\!,\!\sigma')\label{eq:et0} \\
 & =\frac{F^{2}}{\rho_{\text{R}}+\rho_{\text{L}}}(\pi D_\text{eff}t)^{-1/2}+\mathcal{O}\left(t^{-3/2}\right).\label{eq:etaeta}
\end{align}
 Similarly, we obtain from Eqs.~\eqref{eq:Kgen} and \eqref{eq:pump}
\begin{align}
\gamma(t)\! & =\!\sum_{\sigma\sigma'}\!\int\! dxdx'f(x)p(x,\sigma,t|x',\sigma',0)\partial_{x'}\rho_{\text{s}}(x',\sigma')\label{eq:k0}\\
 & =\beta_\text{eff}\frac{F^{2}}{\rho_{\text{R}}+\rho_{\text{L}}}(\pi D_\text{eff}t)^{-1/2}+\mathcal{O}\left(t^{-3/2}\right)~.\label{eq:kt-1}
\end{align}
Remarkably, the long-time regime satisfies an effective FDT $\gamma(t)=\beta_\text{eff}C_\mathcal{F}(t)+\mathcal{O}(t^{-3/2})$. We also note that Eqs~\eqref{eq:st}-\eqref{eq:kt-1} hold in the infinite-system-size limit. For large-but-finite systems, they are complemented by ${\cal O}(L^{-1})$ corrections, as discussed in Sec. III of the Supplemental Material~\cite{SM}.

Equations~\eqref{eq:etaeta} and \eqref{eq:kt-1} immediately show that the asymmetric tracer undergoes anomalous dynamics on long times. Indeed, the noise and friction intensities, defined as $I=\int_{0}^{\infty}dt\CF(t)$
and $\gamma_{\text{T}}=\int_{0}^{\infty}dt\thinspace\gamma(t)$ are infinite, hence leading to an ill-defined diffusivity $D\equiv I/(\gamma_0+\gamma_{\text{T}})^2$. To characterize the anomalous dynamics of the tracer we first  consider its free motion. We  define the tracer's mobility $B(t)$ through $X(t)=\int_{0}^{t}dt'B(t-t')\mathcal{F}(t')$, which leads to
\begin{align}
\left\langle X(t)^{2}\right\rangle _{\text{c}} & =2\int_0^t\!dt_1\!\int_0^{t_1}\!dt_2 B(t_1)B(t_2)C_\mathcal{F}(t_1-t_2)~.
\end{align}
Since we are working in the large $\gamma_0$ limit, $B(t)\sim 1/\gamma_0$~\footnotetext[3]{the diverging behavior of $\gamma_\text{T}$ sets an upper bound on the time-scale for which this approximation holds, as discussed at the end of the Letter.}\cite{Note3}. Using Eq.~\eqref{eq:etaeta} for $C_\mathcal{F}(t)$ then gives Eq.~\eqref{eq:supdiff}, hence implying \emph{superdiffusion}, with
\begin{align}
K & =\frac{4 F^2}{3\rho_{0}\gamma_0^2 \sqrt{\pi D_{\rm eff}}}~.\label{eq:Dta}
\end{align}

In addition to anomalous diffusion, the asymmetric tracer experiences friction that grows with time, as shown by the following towing experiment. Setting a constant velocity $\dot{X}=U$ in Eq.~\eqref{eq:GLE}, the friction exerted by the active particles on the tracer can be measured as $f_\text{fric}(t)\equiv\langle F_\text{tot}\rangle-F$. From Eqs.~\eqref{eq:df} and~\eqref{eq:kt-1}, we get
\begin{align}
f_\text{fric}(t) & = -U\int_{0}^{t}\!dt'\gamma(t')\sim -U\frac{F^{2}}{T_\text{eff}\rho_{0}}\left(\frac{t}{\pi D_\text{eff}}\right)^{1/2},
\end{align}
which yields Eq.~\eqref{eq:fric} with $\Gamma_{\text{T}}=F^{2}(\pi D_\text{eff})^{-1/2} /T_\text{eff}\rho_{0}$.

\emph{Symmetric tracer.}—For a symmetric tracer, $F=0$. Equations~\eqref{eq:etaeta} and \eqref{eq:kt-1}
then imply that $\gamma(t),\ C_\mathcal{F} (t)=\mathcal{O}\left(t^{-3/2}\right)$. In this case, $I$ and $\gamma_{\text{T}}$ remain finite so that $D=I/(\gamma_0+\gamma_{\text{T}})^{2}$ is well defined and Eq.~\eqref{eq:diff} holds. We now present heuristic discussions of $\CF(t)$ and $\gamma(t)$ that account for two important features: their scaling as $t^{-3/2}$ and their sign changes for small tracers. These results can be derived exactly, within the adiabatic limit, for piecewise linear potentials~\cite{OmerFuture}.

Consider a symmetric tracer of length $L_\text{T}$ whose potential is depicted in Fig.~\ref{fig:simpleArg}. While our results can be derived exactly~\cite{OmerFuture}, we present here a simple argument which holds in the limit in which the edges of the tracer have a small width $d$ and small slopes $\pm f_0$. Consider first a single particle located at the left end of the tracer, at $\hat x\simeq 0$, moving in the direction $\hat\sigma$. At long times, the probability distribution of its position $x$ is a Gaussian centered around $\hat \sigma \ell_\text{p}$, of variance $2 D_{\rm eff} t$ (see Fig.~\ref{fig:simpleArg}). The force-force correlation of this particle can be computed as
\begin{equation}\label{eq:pouet}
c(\hat \sigma,t)=    \frac{ f_0^2d}{\sqrt{4 \pi D_\text{eff} t}} \left[e^{-\frac{\ell_\text{p}^2}{4 D_\text{eff} t}}-e^{-\frac{(L_\text{T}-\hat \sigma \ell_\text{p})^2}{4 D_\text{eff} t}}\right]\;,
\end{equation}
as can be inferred from Eq.~\eqref{eq:et0} using $\rho_s(x',\sigma')= \delta(x')\delta_{\sigma',\hat\sigma}$. Note that the factor $d$ comes from the integration over $x$ in Eq.~\eqref{eq:et0}, which also leads to the two exponentials corresponding to $x\simeq 0$ and $x\simeq L_T$, respectively. This amounts to summing the contribution due to particles returning to the left end, such that $f(x)f(x')=f_0^2$, and  that of particles crossing the tracer, such that $f(x)f(x')=-f_0^2$.

Let us return to the case of an active bath of density $\rho_0$. We denote by $m$ the polarization of particles around $x'=0$ so that the local density of particles with orientation $\sigma$ is $\rho_0 \frac{1+\sigma m}2$. The force-force correlation is then obtained from the single-particle result through $\CF(t)=2\rho_0[\frac{1+m}2 c(1,t)+\frac{1-m} 2 c(-1,t)]$, where the factor $2$ stems from the contributions of particles starting at $x'\simeq L_\text{T}$. Expanding the exponentials in Eq.~\eqref{eq:pouet} in the long-time limit, one finds the leading orders to cancel, yielding the $t^{-3/2}$ scaling of $\CF(t)$. Using Eq.~\eqref{eq:st} leads to $m=\mu f_0/v$, which is consistent with the fact that active particles polarize against external potentials~\cite{Enculescu2011}. Straightforward algebra then gives
\begin{align}\label{eq:CFsym}
C_\mathcal{F}(t)  & \!\sim\!\frac{\rho_{0}(f_0 d L_\text{T})^{2}}{4\pi^{1/2}(D_\text{eff}t)^{3/2}} G(\ell_\text{p}/L_\text{T})\;
\end{align}
where $G(y)=1-\frac{2\mu f_0}{v}y$. Importantly, $\CF(t)$ becomes negative when the size of the tracer is small, $L_\text{T}\leq 2 \mu \ell_\text{p} f_0/v$. In the discussion above, we neglected $\mathcal{O}(f_0)$ corrections to the propagator and to the steady-state density due to the edges of the tracer. Including the $f_0$ corrections to all orders confirms the scaling [Eq.~\eqref{eq:CFsym}], to order $\mathcal{O}(d^2)$, albeit with $G(y)=[1-(\frac{2\mu f_0 y}{v})^2]/[1-(\frac{\mu f_0}{v})^2]^2$ (See Sec. IV of the Supplemental Material~\cite{SM}). 
This does not change the leading order estimate for the crossover length $\sim 2\mu f_0\ell_\text{p}/v$. Negative autocorrelations have been reported in other contexts, in~\cite{VanBeijeren1982} and out~\cite{Golestanian2009} of  equilibrium. Here, it is a direct consequence of the polarization against the potential. Setting $m=0$ in the computation above always leads to $\CF(t)>0$.

\begin{figure}
\includegraphics[width=1\columnwidth]{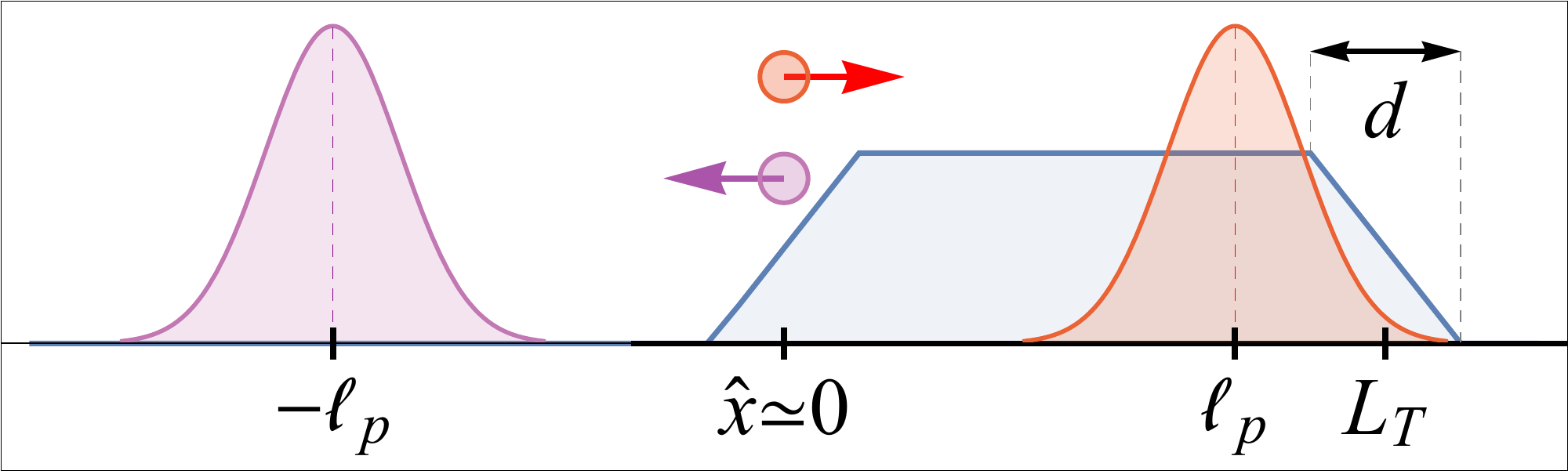}
\caption{Consider a symmetric tracer (blue potential) and an active particle located at its left end at position $\hat{x}$ at $t=0$. The particle is shown in orange and magenta for $\hat{\sigma}=\pm 1$ respectively. At late times, the particle position is distributed as a Gaussian centered around $x_\text{c}=\hat{x}+\hat{\sigma} \ell_\text{p}$. For $\hat{\sigma}=1$, when $\ell_\text{p}\gg L_\text{T}$, the anticorrelation between $f(x')$ and $f(x)$ leads to a negative contribution to $\CF(t)$. Conversely, a $\hat{\sigma}=-1$ particle leads to a positive contribution to $\CF$. Due to the polarization against the potential, $\hat{\sigma}=\pm 1$ occur with different probabilities. This leads to an overall negative $\CF(t)$ for small $L_\text{T}$ and a positive one for large sizes.}\label{fig:simpleArg}
\end{figure}

We now turn to the long-time behavior of $\gamma(t)$. Inserting Eq.~\eqref{eq:st} in Eq.~\eqref{eq:Kgen} leads to  $\gamma=\gamma_{\text{p}}-\gamma_{\text{a}}$, with
\begin{align}
\gamma_{\text{p}}(t-t') & \equiv\beta_\text{eff}\left\langle \mathcal{F}(t)\frac{\mathcal{F}(t')}{1-\left[\frac{\mu}{v}\mathcal{F}(t')\right]^{2}}\right\rangle_\text{c}~,\label{eq:ga0}\\
\gamma_{\text{a}}(t-t') & \equiv\beta_\text{eff}\left\langle \mathcal{F}(t)\frac{\sigma(t')\ell_\text{p}\partial_{x(t')}\mathcal{F}(t')}{1-\sigma(t')\frac{\mu}{v}\mathcal{F}(t')}\right\rangle_\text{c}~.\label{eq:gaa}
\end{align}
The heuristic argument developed above for $\CF(t)$ directly extends to the correlators~\eqref{eq:ga0} and~\eqref{eq:gaa}, showing that $\gamma_\text{a}$ and $\gamma_\text{p}$ both inherit the $t^{-3/2}$ scaling of $\CF(t)$ at long times.  Inspecting Eq.~\eqref{eq:ga0} shows that, to leading order in $f_0$,
\begin{align}\label{eq:FDT}
\gamma_\text{p}(t) \sim \beta_{\rm eff} \CF(t)= \frac{\beta_{\rm eff}\rho_{0}(f_0dL_\text{T})^{2}}{4\pi^{1/2}(D_\text{eff}t)^{3/2}}G(\ell_\text{p}/L_\text{T})\;.  
\end{align}
Equation~\eqref{eq:FDT} is nothing but an effective FDT for the passive tracer. Our results show that the FDT is only expected to hold for small $f_0$ and should be generically violated when $\gamma_a$ is not negligible compared with $\gamma_\text{p}$.

The presence of $\sigma(t')$ in  Eq.~\eqref{eq:gaa} makes the contributions of $\sigma'=\pm 1$ particle add up, instead of canceling, leading to $\gamma_\text{a}(t)>0$ for all $L_\text{T}$ and a long-time scaling $\gamma_\text{a}\sim\mathcal{O}(f_0^3)t^{-3/2}$. Therefore, to leading order in $f_0$, $\gamma \sim  \beta_\text{eff} \CF(t)$. This suggests that $\gamma_\text{T}=\int_0^\infty dt \gamma(t)$ can also change sign and become \emph{negative} for small tracers. Indeed, a perturbative calculation finds that
\begin{align}
    \gamma_\text{T} &\sim \beta_\text{eff}v^{-1}\rho_0(f_0 d)^2\frac{L_\text{T}}{\ell_\text{p}}\left(1-\frac{d^2+6\ell_\text{p}^2}{3dL_\text{T}}\right)~.\label{eq:gT}
\end{align}
The derivations of this result and of the asymptotics of $\gamma_\text{a}$ are not particularly illuminating; they are deferred to Sec. IV of the Supplemental Material~\cite{SM}. Importantly, Eq.~\eqref{eq:gT} implies that when a small symmetric tracer is dragged at velocity $U$, the active bath \emph{enhances} its motion rather than resisting it.

\emph{Adiabatic limit.} Although Eq.~\eqref{eq:GLE} is a common framework to describe a tracer's dynamics, it relies on the assumption that the motion is slow. An important---but rarely debated---question is thus its range of validity. Here, this is set by the requirement that the tracer's response is much slower than the diffusive relaxation of the bath, \emph{i.e.} $\left\langle X(t)\right\rangle,\, \left\langle X^{2}(t)\right\rangle _{\text{c}}^{1/2}\ll(D_{\text{eff}}t)^{1/2}$. For an asymmetric tracer, using $\left\langle X(t)\right\rangle \sim Ft/\gamma_{0}$ and Eq.~\eqref{eq:supdiff}, we find $t\ll \tau_1 \equiv  D_{\text{eff}}(\gamma_{0}/F)^{2}$ and $t\ll \tau_2 \equiv \left(D_{\text{eff}}/K\right)^{2}$. Equation~\eqref{eq:Dta} implies $\tau_1\ll\tau_2$ so that the adiabatic limit holds up to $t\ll\tau_1$. Beyond this timescale, which can be arbitrarily large, an asymmetric tracer in an active bath cannot be described by Eq.~\eqref{eq:GLE}. Considering a finite system of size $L$, the diffusive relaxation time is $t=\tau_{\text{rel}}\sim L^{2}/D_{\text{eff}}$. Thus, the adiabatic limit for an asymmetric tracer in a finite system is valid for $F L \ll D_{\text{eff}}\gamma_0$, which can be achieved by designing the tracer shape to bound $F$ or by using a small enough system. For a symmetric tracer, there is no temporal restriction, and the only requirement is $D\ll D_{\text{eff}}$, which can be fulfilled by setting $\gamma_0\gg(I/D_\text{eff})^{1/2}$. For towing both asymmetric tracers and symmetric tracers at constant velocity $U$, the only requirement is $U\ll D_\text{eff}/L$.

\emph{Conclusion.} 
In this Letter, we have derived the long-time dynamics of a passive tracer in a dilute active bath under the sole assumption of an adiabatic evolution. We have revealed new regimes for both asymmetric and symmetric tracers. First, ratchet effects generically lead to the superdiffusion of asymmetric tracers, which also experience friction that grows with time when they are dragged at constant velocity $U$.  For symmetric tracers, the long-time tail preserves the diffusive behavior, but negative active friction is observed for small tracers. The latter solely follows from the persistent motion of active particles and their polarization by external potentials, a mechanism that differs from previously studied cases with negative mobility~\cite{*[{See, e.g., }] [{ and references therein.}] Cividini2018,Ghosh2014,Maes2020}. We expect the tails for asymmetric and symmetric tracers to become $t^{-d/2}$ and $t^{-(d/2+1)}$ in $d$ dimensions, respectively. This suggests, in two dimensions, that $\left\langle X(t)^{2}\right\rangle _{\text{c}}\sim t\ln t$ for an asymmetric tracer, which remains to be verified. Our results stem from generic features of dry active particles and should thus hold generically. The exponents are expected to be universal, but the transport coefficients can be dressed, for instance, by interactions. 
Moreover, the mechanisms should lead to even richer behaviors for active suspensions in momentum-conserving fluids~\cite{Chen2007,Zaid2011,Thiffeault2015,Kanazawa2020}, or in the presence of phoresis~\cite{Golestanian2009}.

\let\oldaddcontentsline\addcontentsline
\renewcommand{\addcontentsline}[3]{}
\begin{acknowledgments}
We thank Yongjoo Baek, Bernard Derrida, and Xinpeng Xu for many useful discussions. OG and YK are supported by Israel Science Foundation Grant No. 1331/17 and NSF-BSF Grant No. 2016624. JT is supported by the ANR grant THEMA. O. G. also acknowledges support from the Adams Fellowship Program of the Israeli Academy of Sciences and Humanities.
\end{acknowledgments}

\bibliographystyle{apsrev4-2}
\bibliography{references}
\let\addcontentsline\oldaddcontentsline


\pagebreak
\onecolumngrid
\title{Supplemental Material for: "The Anomalous Transport of Tracers in Active Baths"}

\maketitle
\onecolumngrid
\numberwithin{equation}{section}
\numberwithin{figure}{section}
\tableofcontents{}

\section{Simulation details}

\subsection{Setup}

\begin{figure}
\begin{centering}
\includegraphics[width=1\textwidth]{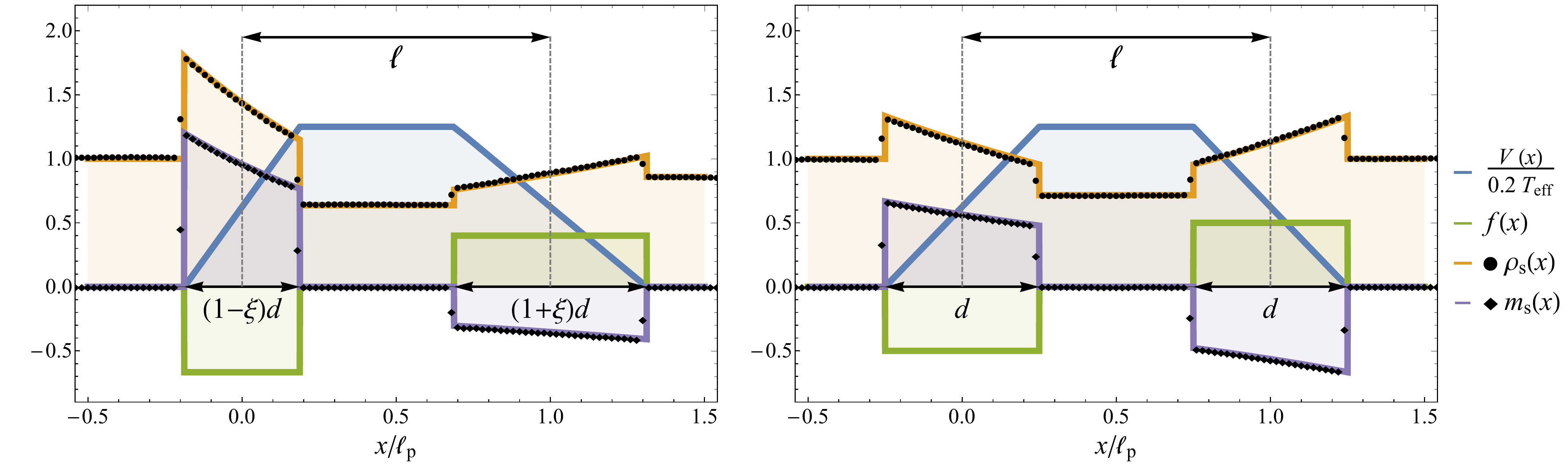}
\par\end{centering}
\caption{\label{fig:Body-potential}Tracer potential $V(x)$, force $f(x)=-\partial_{x}V(x)$, steady-state
density $\rho_{\text{s}}(x)$ and steady-state magnetisation $m_{\text{s}}$ (theory: solid lines; simulation: symbols). Left: asymmteric tracer ($\xi=1/4$); right: symmetric tracer ($\xi=0$). In this figure, $f_0=d=1/2$ and $\rho_\text{L}=\alpha=\mu=v=\ell=1$.}
\end{figure}

We simulate one-dimensional non-interacting run-and-tumble particles
(RTPs) in the presence of a passive tracer. The tracer interacts with
the RTPs via the piecewise linear potential $V(x)$ depicted
in Fig.~\ref{fig:Body-potential}. The potential 
has a total width $L_{\text{T}}=\ell+d$. Its left end has a slope $f_0/(1-\xi)$ and a length $(1-\xi)d$. Its right end has a slope $f_0/(1+\xi)$ and a length $(1+\xi)d$ so that $\xi$ is a measure of the tracer's asymmetry. We choose $f_0$ such that $\mu f_0<v$ and the particles can pass over the obstacle, hence mimicking the ability of active particles to circulate around a finite tracer in a narrow two-dimensional channel. In
all of our simulations, we set $\rho_{\text{L}}=\mu=v=\alpha=1$. All our simulations were ran until a final time $T=10^{4}$.

\subsection{Active particles}

For the time integration of Eq.~(7) we employ an Euler time-stepping: the position of a particle $x_{i+1}$
at time $t_{i+1}=t_{i}+\Delta t$ is given by
\begin{align}
v_{i} & =v\sigma_{i}+\mu f_{i}~,\label{eq:int}\\
x_{i+1} & =x_{i}+v_{i}\Delta t~,\label{eq:int1}
\end{align}
where $\sigma_{i}=\sigma(t_{i})$, $f_{i}=f(x_{i}-X_{i})$, $X_{i}=X(t_{i})$ and $X_i$ is the left end of the tracer at time $t_i$.

The tumbling mechanism is implemented using a continuous-time
Monte Carlo method as follows. At time $t_{0}=0$, a tumbling time
$\tau_{0}$ is drawn from an exponential distribution with mean $2/\alpha$. For all $i>1$, if $\tau_{0}\not\in[t_{i},t_{i+1})$,
the integration steps Eqs.~\eqref{eq:int}-\eqref{eq:int1} are performed
without change. Otherwise, the integration is done up to $\tau_0$, and the next tumbling time $\tau_1=\tau_0+\delta t$ is chosen by sampling $\delta t$ from an exponential distribution with mean $2/\alpha$. The integration of Eqs.~\eqref{eq:int}-\eqref{eq:int1} then continues until $\min(\tau_1,t_{i+1})$ and the above process is repeated. 

A further modification to the time stepping described above is implemented to  account for the transitions between
the different constant-force regions depicted in Fig.~\ref{fig:Body-potential}. We denote the boundaries of these regions by $\{\chi_{j}\}$
and choose the convention $\forall x\in[\chi_{j},\chi_{j+1}),$ $f(\chi_{j})\equiv f(x)$.
Then, if, at a given integration step, $\chi_{j}\in[x_{i},x_{i+1})$, $[t_{i},t_{i+1})$ is partitioned into two intervals $[t_{i},t_{i}+(\chi_{j}-x_{i})/v_{i})$
and $[t_{i}+(\chi_{j}-x_{i})/v_{i},t_{i+1})$ where $v_{i}'=v\sigma_{i}+ \mu f_0(\chi_{j+(\sigma_{i}-1)/2})$. Then, the particles evolve according to Eqs.~\eqref{eq:int}-\eqref{eq:int1}.
If a tumbling event occurs within $[t_{i},t_{i+1})$, the above rule
is applied to each of the two intervals before and after the event. 

\subsection{Tracer}

For the integration of Eq.~(6), we have employed
the midpoint alogrithm, where the tracer velocity $U_{i}=U(t_{i})$
and position $X_{i}=X(t_{i})$ are updated according to 
\begin{align}
U_{i+1} & =F_\text{tot}^i/\gamma_{0}~,\label{eq:Ui}\\
X_{i+1} & =X_{i}+\frac{U_{i}+U_{i+1}}{2}\Delta t~,
\end{align}
where $F_\text{tot}^i=F_\text{tot}(t_i)$. For each $i$, this integration
step is performed after all of the bath particle positions and orientations
have been updated according to the previous section. For this reason, and because active particles are integrated using the Eulerian scheme, the midpoint alogithm does not improve beyond the former. However, we have chosen this algorithm as it provides smoother tracer displacement profiles on short time scales. For the towing simulations at constant speed $U$, Eq.~\eqref{eq:Ui} is replaced by $U_{i}=U$.

The piecewise-linear choice of potential and the implementation of the
two partitioning mechanisms allow integrating Eq.~(7) of the main
text \emph{exactly}, given that the tracer is held fixed. Once the
tracer moves at a finite velocity, the simulation is subject to a finite
accuracy since the tracer position and velocity are updated only after
the bath integration step. Since the simulations are performed in
the adiabatic limit, this accuracy is very high, as seen from the
results presented in the main text.

\subsection{Relaxation to the steady state}

For a given simulation time $T$, we set the system size to $L=v T+L_\text{T}$,
so that there are no finite-size corrections to the propagator: an
active particle that leaves the tracer at time $t=0$ cannot cross
the system and return to it from the other side.

We initiate the simulation at time $t_{\text{i}}=-2T$ with a uniform
distribution of orientations and a static tracer at $X=0$.
Then, the system is let to relax towards its steady state until time
$t=0$, at which the tracer is released and observables are measured
up to time $t=T$. As shown in Fig.~\ref{fig:Body-potential}, this protocole is sufficient for the distribution at $t=0$ to be undistinguishable from the analytical steady-state. For increased performance, in the towing simulations,
the initial distribution was chosen to be the steady-state distribution
$\rho_{\text{s}}(x,\sigma)$ directly. Finally, we verified that all our simulations fall within the adiabatic regime, \emph{i.e.} that $\left\langle X(t)\right\rangle ,\ \left\langle X^{2}(t)\right\rangle _{\text{c}}\ll(D_{\text{eff}}t_{\text{i}})^{1/2}$ is verified.

\subsection{Parameters}

In the free tracer experiments depicted in Fig.~2a of the main text, we set $\xi=0.4$ for the asymmetric tracer. For both tracers we set $\ell=1$, $d=1/2$, $f_0=1/2$ and $\gamma_{0}=10^{3}$, which ensures the validity of the adiabatic limit. To obtain the mean-square displacements, we average the square-displacements over $N=9800$ realizations of the experiment.

In the towing experiments depicted in Figs.~2b\&c, to obtain a higher signal-to-noise ratio, we set $\ell=1$, $d=1/2$, $f_0=1/2$ and $\xi=-0.4749$ for the asymmetric tracer. We average the friction force over $N=1200$ realizations, and apply a temporal moving-average filter of width $100$. For the symmetric tracer, we set $\rho_0=4$, $f=1/40$, $d=\ell$ and vary $\ell$. We average the friction force over $N=10^3-10^4$ realizations. To obtain Fig. 2c, we time-average the resulting force over the last two decades $10^2\leq t\leq10^4$. The relaxation to the asymptotic value of the friction force was verified a posteriori. In all experiments, excluding the towing of a symmetric tracer of size $L_\text{T}=1$, the time step was chosen to be $\Delta t=10^{-2}$. For the former, for increased accuracy, $\Delta t=2.5\times10^{-3}$ was chosen instead.

\subsection{Results for softly interacting active particles}

\begin{figure}
    \begin{centering}
    \includegraphics[width=0.49\textwidth]{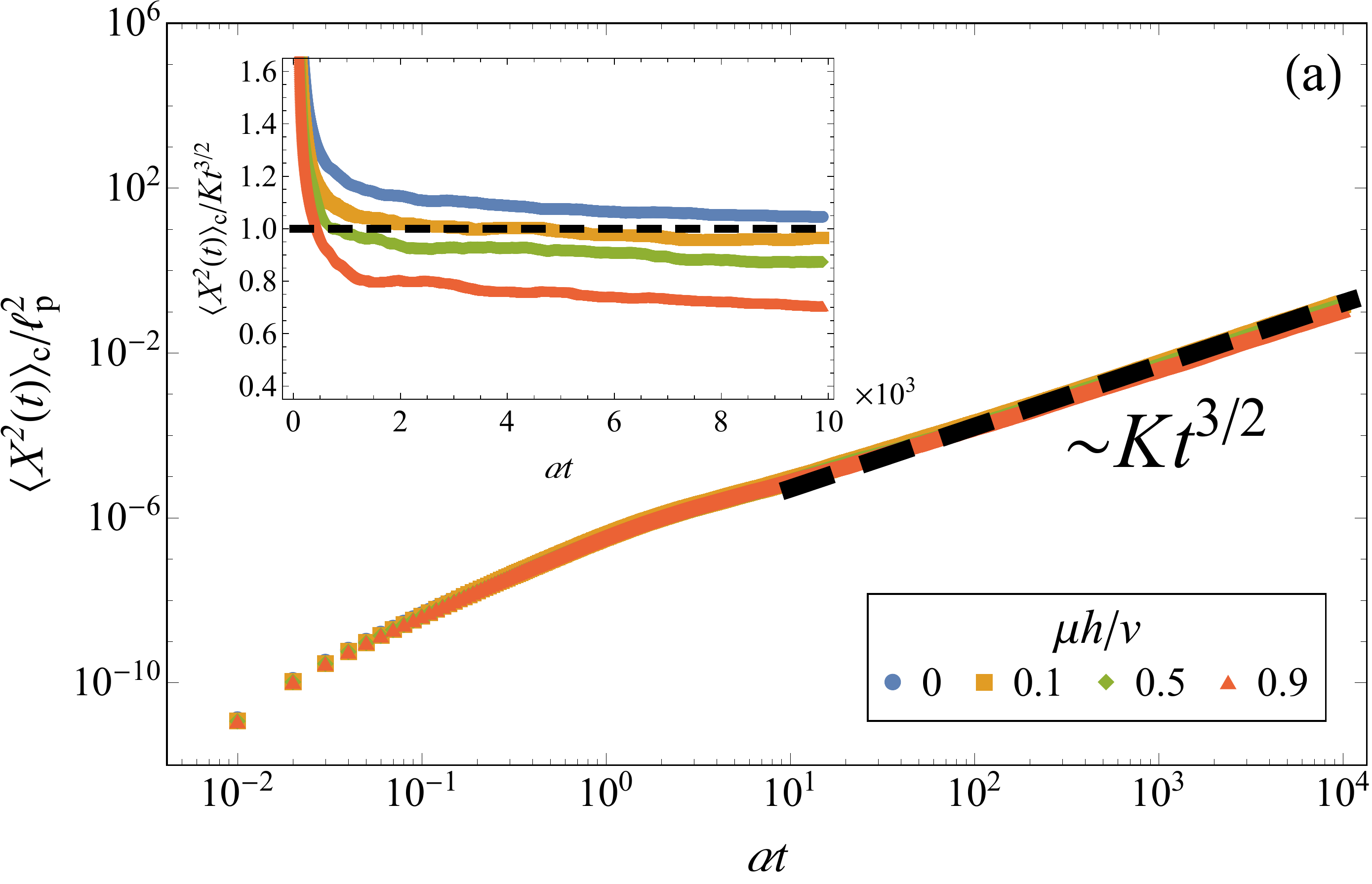}
    \includegraphics[width=0.49\textwidth]{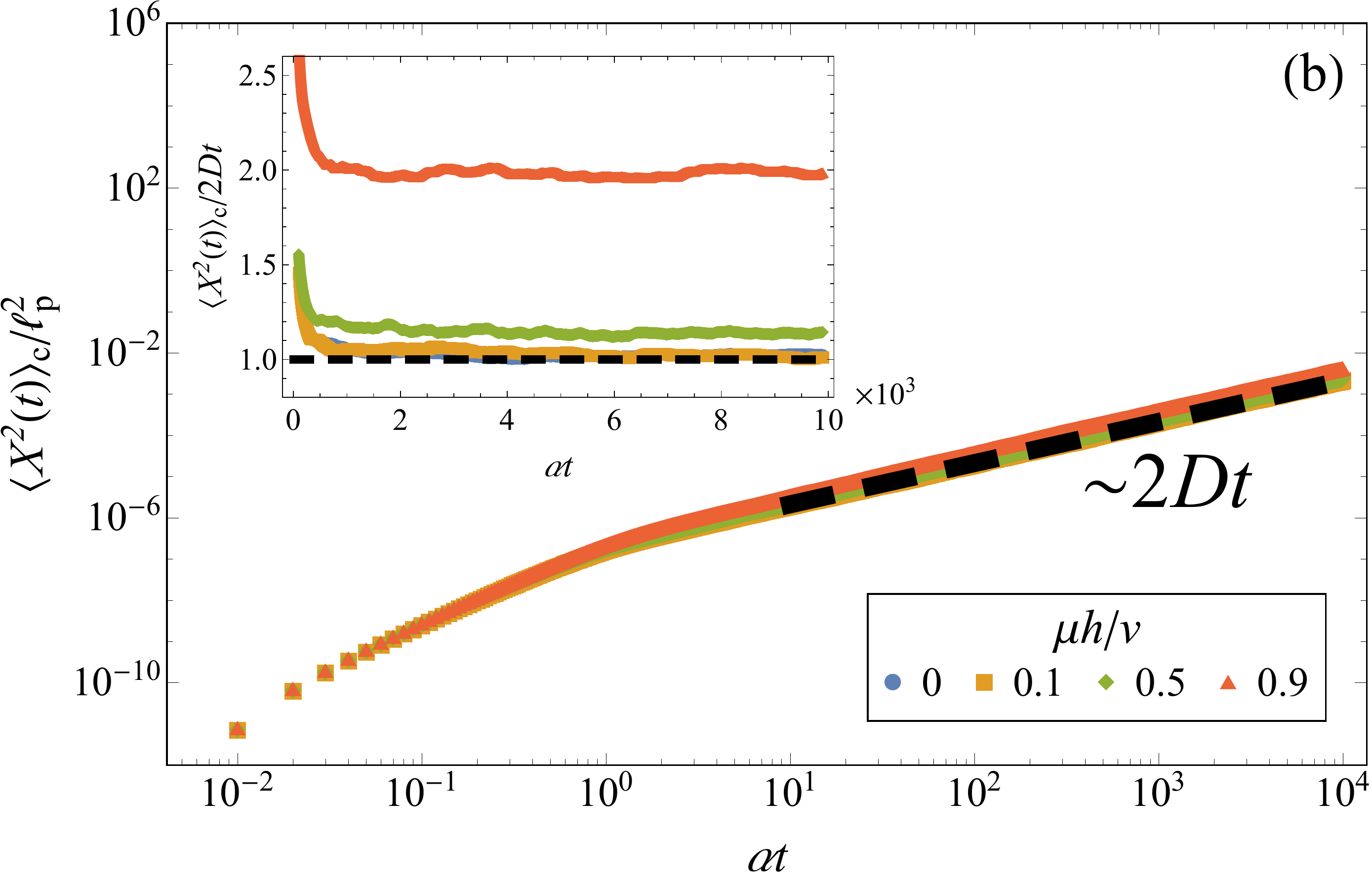}
    \par\end{centering}
    \caption{Simulation results for a tracer in a bath of softly interacting active particles (symbols) compared with our theoretical predictions for the long-time limit of the non-interacting case (dashed black lines). The exponents are unaltered, revealing the universal nature of our predictions. As expected, the prefactor is renormalized by interactions, as illustrated in the insets. (a) Asymmetric tracer and (b) symmetric tracer. Insets: MSDs divided by the theoretical predictions (a moving average filter of width $100~\alpha^{-1}$ is applied to the data). $h$ is the amplitude of the interparticle force (see text).}
    \label{fig:interactions}
\end{figure}

Here we present numerical results for a bath of pairwise interacting active particles. We consider the interaction potential $V_\text{int}(x)=h(a-|x|)\Theta(a-|x|)$ and set $a=\ell_\text{p}/10$. We reproduce Fig.~2a of the main text for various values of $h$ (see Fig.~\ref{fig:interactions}). The asymptotic behaviour on long times agrees well with our theoretical predictions: while the exponents are unaffected by the interactions, the prefactors are renormalized by the interactions. The exponents of the MSDs are universal as expected through the arguments of the main text.

\section{Systematic derivation of the long-time limit of the propagator $p(x,\sigma,t|x',\sigma',0)$}
\label{sec:longtime}

In the main text, we present a self-contained heuristic derivation of the long-time limit of the propagator given by Eq.~(13). For sake of completeness, we present here its systematic derivation. Our method is a perturbation theory for which the small parameter is $1/t$ as $t\rightarrow\infty$.

To proceed, we note that the stochastic differential equation~(7) of the main text is equivalent to the master equation for the time-evolution of the probability densities of finding a right-moving and left-moving RTP at position $x$ and time $t$, denoted by $p_{+1}(x,t)$ and $p_{-1}(x,t)$, respectively. The equation reads
\begin{align}
\partial_{t}p_{\sigma} & =-\partial_{x}\left[\left(\sigma v+\mu f\right)p_{\sigma}\right]-\frac{\alpha}{2}\left(p_{\sigma}-p_{-\sigma}\right)~,\label{eq:rtpt}
\end{align}
where the force is given by $f(x)=-\partial_{x}V(x)$. The propagator $p(x,\sigma,t|x',\sigma',0)$, for which we
use the shorthand notation $p_{\sigma}(x,t)$, is the solution of Eq.~\eqref{eq:rtpt} for the initial condition
$p_{\sigma}(x,0)=\delta_{\sigma\sigma'}\delta(x-x')$. To compute $p_{\sigma}(x,t)$, we note that the dynamics~\eqref{eq:rtpt}, which couple $\sigma$ and $-\sigma$, can be decoupled by applying $\partial_{t}$ to both sides of the equation, which leads to
\begin{align}
0 & =\left(\partial_{t}^{2}+\alpha\partial_{t}+\partial_{x}J_{\sigma}\right)p_{\sigma}~,\label{eq:Mt0}\\
J_{\sigma}(x,\partial_{t}) & \equiv-\left[v^{2}-\left(\mu f\right)^{2}\right]\partial_{x}+\mu\left[f\left(\alpha+2\partial_{t}\right)-\left(\sigma v-\mu f\right)f'\right]~.
\end{align}
For clarity, we rescale the system as $t\rightarrow\alpha^{-1}t$,
$x\rightarrow \ell_{\text{p}}x$ and $V\rightarrow T_{\text{eff}}V$, which is equivalent to setting $\alpha=\mu=v=1$. The result is
\begin{align}
0 & =\left(\partial_{t}^{2}+\partial_{t}+\partial_{x}J_{\sigma}\right)p_{\sigma}~,\label{eq:Mt}\\
J_{\sigma}(x,\partial_{t}) & =-\left(1-f^{2}\right)\partial_{x}+\left[f\left(1+2\partial_{t}\right)-\left(\sigma-f\right)f'\right]~.\label{eq:Mt2}
\end{align}
We proceed by taking the Laplace transform of Eq.~\eqref{eq:Mt} with
respect to $t$, which yields
\begin{align}
S_{\sigma\sigma'}\delta(x-x') & =\left[s^{2}+s+\partial_{x}J_{\sigma}(x,s)\right]p_{\sigma}(x,s)~,\label{eq:FPs}\\
S_{\sigma\sigma'}(x) & \equiv\left[s-\partial_{x}\left(\sigma- f(x)\right)\right]\delta_{\sigma\sigma'}+\frac{1}{2}~,\label{eq:FP1}
\end{align}
where $p_{\sigma}(x,s)\equiv\int_{0}^{\infty}dte^{-st}p_\sigma(x,t)$ denotes
the Laplace transform of $p_\sigma(x,t)$ and $\partial_x$ acts on everything to its right. Next, we decompose the current and source operators into force-independent and force-dependent components,
\begin{align}
    J_\sigma &= -\partial_x+J_\sigma^f~,\label{eq:Jsdec}\\
    S_{\sigma\sigma'} &= S_{\sigma\sigma'}^0+\partial_xf(x)\delta_{\sigma\sigma'}~,\\
    J_{\sigma}^{f}(x,s) & \equiv f^2\partial_{x}+\left[f\left(1+2s\right)-\left(\sigma- f\right)f'\right]~,\label{eq:Jf}\\
    S_{\sigma\sigma'}^0(x) & \equiv\left(s-\sigma\partial_{x}\right)\delta_{\sigma\sigma'}+\frac{1}{2}~,
\end{align}
which allows rewriting Eq.~\eqref{eq:FPs} as
\begin{align}
\left(M_{0}+\partial_{x}J_{\sigma}^{f}\right)p_{\sigma} & =\left[S_{\sigma\sigma'}^0+\partial_x f(x)\delta_{\sigma\sigma'}\right]\delta(x-x')~,\label{eq:intermediaire}\\
M_{0} & \equiv s^{2}+s-\partial_{x}^{2}~.\label{eq:M0}
\end{align}

Before solving Eq.~\eqref{eq:intermediaire} in the general, it is instructive to consider the case $f(x)=0$ to see how the propagator can be expanded at large times, i.e. for small $s$. Since $f=0$, then one also has that $J_{\sigma}^{f}=0$ and the solution of Eq.~\eqref{eq:intermediaire} is
\begin{align}
p_{\sigma}^0(x,s)&=M_{0}^{-1}S_{\sigma\sigma'}^{0}\delta(x-x')\\
& =\left\{\frac{1}{2}\left[\sigma\text{sgn}(x-x')+\frac{s+\frac{1}{2}}{(s^{2}+s)^{1/2}}\right]\delta_{\sigma\sigma'}+\frac{1}{4(s^{2}+s)^{1/2}}\delta_{\sigma,-\sigma'}\right\}e^{-(s^{2}+s)^{1/2}\left|x-x'\right|}~,\label{eq:fs}
\end{align}
where $M_{0}^{-1}=\int dyG_{0}(x-y)$ and $G_{0}(x-x')$ is the Green's
function of $M_{0}$, \emph{i.e.
\begin{align}
M_{0}(x)G_{0}(x-x') & =\delta(x-x')~.
\end{align}
}The solution to this equation is
\begin{align}
G_{0}(x-x') & =\frac{1}{2(s^{2}+s)^{1/2}}e^{-(s^{2}+s)^{1/2}\left|x-x'\right|}\label{eq:G0sol}\\
 & =\frac{1}{2(s^{2}+s)^{1/2}}\sum_{n=0}^{\infty}\frac{(-\left|x-x'\right|)^{n}}{n!}(s^{2}+s)^{n/2}~.\label{eq:g0exp}
\end{align}
Expanding in the limit $s\ll1$, Eq.~\eqref{eq:g0exp} becomes an expansion
in powers of $s^{1/2}$. 

Using the insight gained from the $f=0$ case, we look for $p_{\sigma}$ in the $f(x)\neq0$ case as a series
\begin{align}
p_{\sigma} & =\frac{1}{s^{1/2}}\sum_{n=0}^{\infty}A_{n}(x,\sigma|x',\sigma')s^{n/2}~,\label{eq:exp-1}
\end{align}
where the $A_n$ need to be determined. In real time, Eq.~\eqref{eq:exp-1} 
provides the long-time expansion
\begin{align}
p_{\sigma}(x,t) & =t^{-1/2}\sum_{n=0}^{\infty}\frac{A_{2n}(x,\sigma|x',\sigma')}{\Gamma(\frac{1}{2}-2n)}t^{-n}~.
\end{align}
Inserting Eq.~\eqref{eq:exp-1} into Eq.~\eqref{eq:FPs}, we arrive
at the hierarchy
\begin{align}
\partial_{x}\left[J_{\sigma}(x,0)A_{0}(x,\sigma)\right] & =0~,\label{eq:a0}\\
\partial_{x}\left[J_{\sigma}(x,0)A_{1}(x,\sigma)\right] & =S_{\sigma\sigma'}|_{s=0}\delta(x-x')~,\label{eq:a1}\\
\left[1+2\partial_xf(x)\right]A_{0}(x,\sigma)+\partial_{x}\left[J_{\sigma}(x,0)A_{2}(x,\sigma)\right] & =0~,\label{eq:a2}\\
\vdots\nonumber 
\end{align}
This allows us to solve for $p_{\sigma}$ perturbatively in the limit
$s\ll1$. 

Note that both the steady-state $\rho_\text{s}(x,\sigma)$ and $A_0$ satisfy $\partial_{x}(J_{\sigma}\rho_\text{s})=\partial_{x}(J_{\sigma}A_0)=0$. It is thus tempting to look for $A_0$ of the form
\begin{align}
    A_{0}(x,\sigma|x',\sigma')&=N_{\sigma\sigma'}(x')\rho_{\text{s}}(x,\sigma)~,\label{eq:A0}
\end{align}
where $N_{\sigma\sigma'}(x')$ is \emph{a priori} unknown. We now turn to show that this is indeed the case and to determine $N_{\sigma\sigma'}$. 

First note that the steady-state solution is current free so that, using the notations of Eqs.~\eqref{eq:Mt} and~\eqref{eq:Mt2}, $J_\sigma(x,0)\rho_\text{s}(x,\sigma)=0$. 
Then, Eq.~\eqref{eq:A0} is equivalent to $J_{\sigma}(x,0)A_{0}(x,\sigma)=0$ since $J_{\sigma}(x,0)$ is a first-order linear differential operator in $x$:
by uniqueness of the solution, if both $A_0$ and $\rho_\text{s}$ are in the kernel of $J_\sigma$, then $A_{0}(x,\sigma|x',\sigma')=N_{\sigma\sigma'}(x')\rho_\text{s}(x,\sigma)$
for some $N_{\sigma\sigma'}(x')$.

To show that $J_{\sigma}(x,0)A_{0}(x,\sigma)=0$, we proceed as follows. We use Eq.~\eqref{eq:intermediaire} to relate the propagators with the tracer, $p_\sigma$, and without the tracer, $p_\sigma^0$, and then expand their relationship for small $s$.
Multiplying both sides of Eq.~\eqref{eq:intermediaire}
by $M_{0}^{-1}$, we get
\begin{align}
\left(1+M_{0}^{-1}\partial_{x}J_{\sigma}^{f}\right)p_{\sigma} & =M_{0}^{-1}\left(S_{\sigma\sigma'}^{0}+\partial_{x}f\delta_{\sigma\sigma'}\right)\delta(x-x')=p_{\sigma}^{0}+M_{0}^{-1}\partial_{x}f\delta_{\sigma\sigma'}\delta(x-x')~,\label{eq:fpinv}
\end{align}
with $p_\sigma^0$ given in Eq.~\eqref{eq:fs} and $M_0$ defined in Eq.~\eqref{eq:M0}. Substituting $M_{0}^{-1}=\int dyG_{0}(x-y)$ into Eq.~\eqref{eq:fpinv} then yields
\begin{align}
p_{\sigma}^{0} & =p_{\sigma}+\int dyG_{0}(x-y)\partial_{y}\left[J_{\sigma}^{f}p_{\sigma}- f\delta_{\sigma\sigma'}\delta(y-x')\right]\\
 & =p_{\sigma}+\int dy\partial_{x}G_{0}(x-y)\left[J_{\sigma}^{f}p_{\sigma}- f\delta_{\sigma\sigma'}\delta(y-x')\right]~,\label{eq:Mint-1}
\end{align}
where, in the second equality, we have integrated by parts and used
the spatial symmetry of $G_{0}$. Next, we expand for small $s$.
From Eqs.~\eqref{eq:exp-1} and \eqref{eq:g0exp} we have that
\begin{align}
p_{\sigma} & =A_{0}s^{-1/2}+\mathcal{O}(s^{0})~,\\
p_{\sigma}^{0} & =\frac{1}{4}s^{-1/2}+\mathcal{O}(s^{0})~,\\
G_{0}(x-y) & =\frac{1}{2}s^{-1/2}+\mathcal{O}(s^{0})\label{eq:g0}~,\\
\partial_{x}G_{0}(x-y) & =-\frac{1}{2}\text{sgn}(x-y)+\frac{1}{2}(x-y)s^{1/2}+\mathcal{O}(s)~.\label{eq:dxg0}
\end{align}
Substituting these into Eq.~\eqref{eq:Mint-1} and equating coefficients at order $s^{-1/2}$ gives
\begin{align}
\left[-\int^{x}dyJ_{\sigma}(y,0)+\frac{1}{2}\int dyJ_{\sigma}^{f}(y,0)\right]A_{0}(y,\sigma)&=A_0(x,\sigma)+\left(-\int^{x}+\frac{1}{2}\int\right)dyJ_{\sigma}^{f}(y,0)A_{0}(y,\sigma)\label{eq:int-11}\\ & =\frac{1}{4}~,\label{eq:int-1}
\end{align}
where, in the first equality, we have used Eq.~\eqref{eq:Jsdec}.
Differentiating Eq.~\eqref{eq:int-1} with respect to $x$ then yields
$J_{\sigma}(x,0)A_{0}(x,\sigma)=0$. 

 To determine $N_{\sigma\sigma'}(x')$,
we insert the solution into Eq.~\eqref{eq:int-11}. Using $J_{\sigma}(x,0)A_{0}(x,\sigma)=0$
and Eqs.~\eqref{eq:Jsdec} and~\eqref{eq:A0}, we get
\begin{align}
    J_{\sigma}^{f}(x,0)A_{0}(x,\sigma|x',\sigma')&=\partial_{x}A_{0}(x,\sigma|x',\sigma')=N_{\sigma\sigma'}(x')\partial_{x}\rho_{s}(x,\sigma)~.
\end{align}
Therefore, 
\begin{align}
\int dyJ_{\sigma}^{f}(y,0)A_{0}(y,\sigma|x',\sigma') & =N_{\sigma\sigma'}(x')\int dy\partial_{y}\rho_\text{s}(y,\sigma)=\frac{1}{2}N_{\sigma\sigma'}(x')\left(\rho_\text{R}-\rho_\text{L}\right)~,
\end{align}
where we have used $\rho_\text{s}(x,\sigma)=\rho_\text{s}(x)/2$ outside the tracer.
Finally, Eqs.~\eqref{eq:int-11}-\eqref{eq:int-1} lead to
\begin{align}
N_{\sigma\sigma'}(x')\left[\rho_\text{s}(x,\sigma)-\int^{x}dyJ_\sigma^{f}(y,0)\rho_\text{s}(y,\sigma)\right]+\frac{1}{4}N_{\sigma\sigma'}(x')\left(\rho_\text{R}-\rho_\text{L}\right) & =\frac{1}{4}~.
\end{align}
Choosing any value of $x$ left of the tracer, for which $J_\sigma^f(y,0)=0$ for any $y\leq x$, then yields
\begin{align}
N_{\sigma\sigma'}(x') & =\frac{1}{\rho_\text{R}+\rho_\text{L}}~,
\end{align}
All in all, recalling that $p_{\sigma}(x,t)=p(x,\sigma,t|x',\sigma',0)$ and  restoring dimensions,
we get
\begin{align}
p(x,\sigma,t|x',\sigma',0) & =\frac{\rho_{\text{s}}(x,\sigma)}{\rho_{\text{R}}+\rho_{\text{L}}}(\pi D_{\text{eff}}t)^{-1/2}+\mathcal{O}\left(t^{-3/2}\right)~,\label{eq:propg}
\end{align}
which coincides with the heuristic result Eq.~(13) of the main text.

\section{Finite-size effects}
In Section~\ref{sec:FSSS}, we first review for completeness the steady-state solution of non-interacting RTPs subject to a potential $V(x)$, following~\cite{Kitahara1979,Angelani2011,Solon2015PressureFluids}. This corresponds to Eq. (11) of the main text. We show the solution to apply both to periodic and closed systems, up to finite-size corrections that scale as $\mathcal{O}(L^{-1})$. 

Then, in Section~\ref{sec:FSCFG}, we bound the finite-size corrections to $C_\mathcal{F}(t)$ and $\gamma(t)$ and show that Eqs.~(14) and (16) of the main text hold, up to corrections of order $\mathcal{O}(L^{-1})$. In the process, we show that the infinite-system-size limit can be obtained equivalently using either closed or periodic systems. This allows us to conduct our simulations using periodic boundary conditions. Throughout our discussion, we make the equations dimensionless by rescaling $t\rightarrow\alpha^{-1}t$,
$x\rightarrow \ell_{\text{p}}x$ and $V\rightarrow T_{\text{eff}}V$,
which is equivalent to setting $\alpha=\mu=v=1$. 

\subsection{Steady-state solution}
\label{sec:FSSS}
Using the notations introduced at the beginning of Section~\ref{sec:longtime}, the bath density and magnetization are respectively given by $\rho(x,t)=p_{+1}(x,t)+p_{-1}(x,t)$ and $m(x,t)=p_{+1}(x,t)-p_{-1}(x,t)$, up to a common normalization factor that accounts for multiple non-interacting particles. Using Eq.~\eqref{eq:rtpt}, one arrives at the dynamics
\begin{align}
\partial_t \rho & =-\partial_{x}(vm+\mu\rho f)~,\label{eq:rhoswithDim}\\
\partial_t m & =-\partial_{x}(v\rho+\mu m f)-\alpha m~.\label{eq:mswithDim}
\end{align}
In the rescaled variables, the dynamics reduces to
\begin{align}
\partial_t \rho & =-\partial_{x}(m+\rho f)~,\label{eq:rhot}\\
\partial_t m & =-\partial_{x}(\rho+m f)-m~.\label{eq:mt}
\end{align}
The equations for the steady-state density $\rho_{\text{s}}(x)$ and magnetization $m_{\text{s}}(x)$ are then
\begin{align}
0 & =-\partial_{x}(m_{\text{s}}+\rho_{\text{s}}f)~,\label{eq:rhos}\\
0 & =-\partial_{x}(\rho_{\text{s}}+m_{\text{s}}f)-m_\text{s}~.\label{eq:ms}
\end{align}
From Eq.~\eqref{eq:rhos}, we obtain the steady-state current
\begin{align}
J\equiv m_{\text{s}}+\rho_{\text{s}}f & =\text{const}~.\label{eq:const}
\end{align}
Combining Eqs.~\eqref{eq:ms}
and \eqref{eq:const} provides the steady-state conditions~\cite{Angelani2011}
\begin{align}
    \big\{-\partial_x\left[1-f^2(x)\right]+f(x)\big\}\rho_\text{s}(x)&=J\left[1+\partial_xf(x)\right]~,\label{eq:steq}\\
    m_\text{s}(x)=J-\rho_\text{s}f(x)~,\label{eq:mst}
\end{align}
where $\partial_x$ operates on $\left[1-f^2(x)\right]\rho_\text{s}(x)$. Using Eq.~\eqref{eq:mst}, one can express the full steady-state distribution $\rho_\text{s}(x,\sigma)$ using $\rho_\text{s}(x)$ and $J$ as
\begin{align}
    \rho_\text{s}(x,\sigma)&=\frac{\rho_\text{s}(x)+\sigma m_\text{s}(x)}{2}=\frac{1-\sigma f(x)}{2}\rho_\text{s}(x)+\frac{\sigma}{2}J~.\label{eq:full}
\end{align}
To determine $\rho_\text{s}(x)$ and $J$, one now has to solve Eq.~\eqref{eq:steq} with the appropriate boundary conditions.

In a closed system with boundaries at $\pm L/2$, the current $J$ vanishes for any $L$. In particular, $J=0$ for $L\rightarrow\infty$. The solution to Eq.~\eqref{eq:steq} is then
\begin{align}
\rho_\text{s}(x) & =\frac{\rho_\text{L}}{1-f^2(x)}\exp{\left[\int_{-L}^x dy \frac{f(y)}{1-f^2(y)}\right]}~,\label{eq:stsolrho}
\end{align}
where $\rho_\text{L}$ is the density left of the tracer. Using Eq.~\eqref{eq:full} and~\eqref{eq:mst}, one obtains
\begin{align}
\rho_\text{s}(x,\sigma) & =\frac{\frac{1}{2}\rho_\text{L}}{1+\sigma f(x)}\exp{\left[\int_{-L}^x dy \frac{f(y)}{1-f^2(y)}\right]}~.\label{eq:stsol}
\end{align}
This validates the current-free solution in Eq.~(11) for infinite or finite and closed systems.

For periodic systems, $J\neq0$, and the validity of Eq.~(11) as a large-$L$ limit of the solution on a ring should be justified. The general, unnormalized, solution $\rho_{\text{s}}^{\text{per}}(x)$ to Eq.~\eqref{eq:steq} on a ring reads
\begin{align}
\rho_{\text{s}}^{\text{per}}(x) & =\frac{e^{\int_{-L/2}^{x}dy\frac{f(y)}{1- f^2(y)}}}{1- f^2(x)}\left\{ \rho_{\text{L}}-J\left[f(x)e^{-\int_{-L/2}^{x}dy\frac{f(y)}{1- f^2(y)}}+\int_{-L/2}^{x}dy\frac{e^{-\int_{-L/2}^{y}dz\frac{f(z)}{1- f^2(z)}}}{1- f^2(y)}\right]\right\}~.\label{eq:per}
\end{align}
Using the periodic boundary condition $\rho_{\text{s}}^{\text{per}}(-L/2)=\rho_{\text{s}}^{\text{per}}(L/2)$
yields
\begin{align}
J & =\rho_{\text{L}}\frac{e^{\int_{-L/2}^{L/2}dx\frac{f(x)}{1- f^2(x)}}-1}{\int_{-L/2}^{L/2}dx\frac{e^{\int_{x}^{L/2}dy\frac{f(y)}{1- f^2(y)}}}{1- f^2(x)}}=\rho_{\text{L}}\frac{e^{\int_{-L/2}^{L/2}dx\frac{f(x)}{1- f^2(x)}}-1}{\frac{L-L_{\text{T}}}{2}\left[e^{\int_{-L/2}^{L/2}dx\frac{f(x)}{1- f^2(x)}}+1\right]+L_{\text{T}}I_{0}}~,
\end{align}
where we use the definitions of the main text and introduce
\begin{align}
I_{0} & \equiv\int_{0}^{L_\text{T}}\frac{dx}{L_{\text{T}}}\frac{e^{\int_{y}^{x_{\text{R}}}dy\frac{f(y)}{1- f^2(y)}}}{1- f^2(x)}~.
\end{align}
For $L\gg1$, we obtain the familiar result
\begin{align}
J & =-\frac{F}{L}\left[1+\mathcal{O}\left(\frac{L_{\text{T}}}{L}\right)\right]~,\label{eq:Jfinite}
\end{align}
where $F=-(\rho_\text{R}-\rho_\text{L})$. Expectedly, Eq.~\eqref{eq:Jfinite} shows that the current vanishes as $L\rightarrow\infty$. In this limit, we obtain
\begin{align}
\rho_{\text{s}}^{\text{per}}(x) & =\left(1+\frac{F}{2\rho_{0}}\right)\rho_{\text{s}}(x)~,
\end{align}
where $\rho_0=(\rho_\text{R}+\rho_\text{L})/2$. Therefore, the infinite-size limit of solutions in closed and periodic
systems differs only by an overall multiplicative constant. Rescaling the density as
\begin{align}
\rho_{0} & \rightarrow\frac{\rho_{0}}{1+\frac{F}{2\rho_{0}}}~,\label{eq:rescale}
\end{align}
leads to the same density profile as in the closed case and thus validates Eq.~(11) for the periodic case as well. Note that,
under this rescaling, $\rho_{\text{L}}\rightarrow\rho_{0}$.

\if{\subsection{Propagator}

Here we discuss finite-size corrections to the propagator $p(x,\sigma,t|x',\sigma',0)$. It is the probability to find a particle at position and orientation $(x,\sigma)$ at time $t$, given the deterministic initial condition $p(x,\sigma,0|x',\sigma',0)=\delta_{\sigma\sigma'}\delta(x-x')$. As $t$ increases, the propagation of probability occurs at two different rates. First,
there is a ballistic propagation `front' which is exponentially
suppressed by a factor $\sim e^{-t}$ due to the tumbling. This provides a finite-size correction
$\sim e^{-L}$ to the propagator, which can be neglected for large $L$ values and on long time scales.
Second, there is a diffusive propagation front, which
propagates until the system boundaries are reached. Therefore, the propagator
has no significant finite-size corrections for $t\ll L^{2}$, which sets an upper bound on the maximal time $T$ under which finite-size effects can be safely neglected.}\fi

\subsection{Friction kernel and force-force correlation}
\label{sec:FSCFG}

To conclude the section, we show that the finite-size corrections to Eqs.~(14) and (16) are $\mathcal{O}(L^{-1})$. We first note that we can use the expression~(13) for the propagator, which is derived in an infinite system, as long as the time is such that $t \ll L^2$ so that the boundaries of the system are not reached by RTPs that interact with the obstacle at $t=0$. 

We now want to show that the measurement of the force-force correlations in periodic systems, given by
\begin{align}
    C_{\mathcal{F}}^{\text{per}}(t) & = \sum_{\sigma\sigma'}\int dxdx'f(x)p(x,\sigma,t|x',\sigma',0)f(x')\rho_{\text{s}}^{\text{per}}(x',\sigma')~,
\end{align}    
indeed yields, to leading order in $L^{-1}$, $C_{\mathcal{F}}(t)$ as predicted by Eq.~(14). We first note that, since $f(x)$ has compact support, $C_{\mathcal{F}}^{\text{per}}(t)$ remains finite as $L\rightarrow\infty$ irrespective of the boundary conditions. Then, up to the rescaling given in Eq.~\eqref{eq:rescale},  $\rho_{\text{s}}(x',\sigma')$ and $\rho^{\rm per}_{\text{s}}(x',\sigma')$ coincide up to corrections of order ${\cal O}(L^{-1})$. The force-force correlations predicted by Eq. (14) in the main text thus coincide with those measured in periodic systems, up to the rescaling~\eqref{eq:rescale} and corrections of order ${\cal O}(L^{-1})$.

We now turn to show that the same holds for the friction
\begin{align}
    \gamma^{\text{per}}(t) & =\sum_{\sigma\sigma'}\int dxdx'f(x)p(x,\sigma,t|x',\sigma',0)\partial_{x'}\rho_{\text{s}}^{\text{per}}(x',\sigma')~.\label{eq:gammaper}
\end{align}
In contrast to $C_{\mathcal{F}}^{\text{per}}$, the finiteness of $\gamma^{\text{per}}(t)$ should be verified for large periodic systems because of the integral over $x'$ in Eq.~\eqref{eq:gammaper}. Dividing the integration domain into the region inside the tracer
$\mathcal{T}=\{0<x<L_\text{T}\}$ and outside of the tracer $\mathcal{T}^{\text{c}}$,
we obtain
\begin{align}
\gamma^{\text{per}}(t) & =\sum_{\sigma\sigma'}\int dx\left(\int_{\mathcal{T}}dx'+\int_{\mathcal{T}^{\text{c}}}dx'\right)f(x)p(x,\sigma,t|x',\sigma',0)\partial_{x'}\rho_{\text{s}}^{\text{per}}(x',\sigma')\nonumber \\
 & =\sum_{\sigma\sigma'}\int dx\int_{\mathcal{T}}dx'f(x)p(x,\sigma,t|x',\sigma',0)\partial_{x'}\rho_{\text{s}}^{\text{per}}(x',\sigma')-\frac{J}{2}\sum_{\sigma\sigma'}\int dx\int_{\mathcal{T}^{\text{c}}}dx'f(x)p(x,\sigma,t|x',\sigma',0)~,\label{eq:gp}
\end{align}
where, in the second equality, we used Eqs.~\eqref{eq:steq} and~\eqref{eq:full}, which give $\partial_{x'}\rho_{\text{s}}^{\text{per}}(x',\sigma')=\partial_{x'}\rho_{\text{s}}^{\text{per}}(x')/2=-J/2$
outside the tracer. Since $J\sim - F/L$ (see Eq.~\eqref{eq:Jfinite}), one finds that
\begin{align}
-\frac{J}{2}\sum_{\sigma\sigma'}\int dx\int_{\mathcal{T}^{\text{c}}}dx'f(x)p(x,\sigma,t|x',\sigma',0) & =
\frac{1}{2}F\int dxf(x)p_\text{U}(x,t)+\mathcal{O}\left(\frac{L_\text{T}}{L}\right)~,
\end{align}
where the probability density $p_\text{U}(x,t)\equiv\int dx'p(x,t|x',0)\cdot L^{-1}$ is the propagation of an initially uniform distribution $p_\text{U}(x,0)=L^{-1}$. Because $|f(x)|<1$, every point in the system is accessible by the dynamics. This implies that $p_\text{U}(x,t)$ is spread over the entire system with a non-negligible density at each point. In conjunction with normalization of probability, $\int dx\ p_\text{U}(x,t)=1$, it implies that $p_\text{U}(x,t)\sim L^{-1}$. Since $f(x)=0$ outside of the tracer, it follows that $F\int dxf(x)p_\text{U}(x,t)/2=\mathcal{O}(L_\text{T}/L)$. We conclude that $\gamma^{\text{per}}(t)$ remains finite
as $L\rightarrow\infty$, and can be obtained from $\gamma(t)$ by rescaling the density according to Eq.~\eqref{eq:rescale}. As claimed, the leading order correction to Eq.~(16) is $\mathcal{O}(L^{-1})$.

\section{Perturbative analysis for symmetric tracers}

\begin{figure}
\begin{centering}
\includegraphics[width=0.55\textwidth]{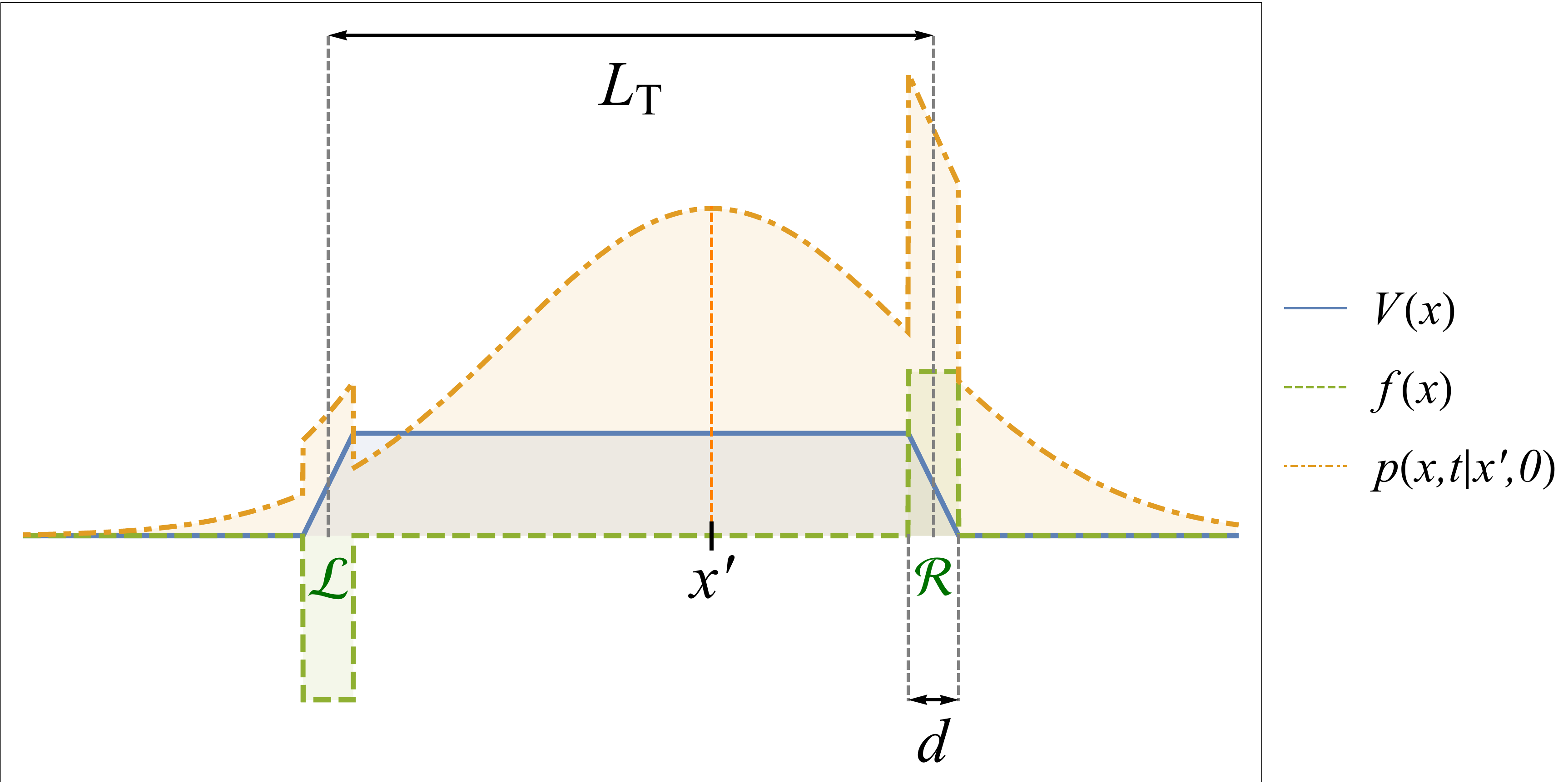}
\par\end{centering}
\caption{\label{fig:pert}Schematic description of the influence of the tracer (blue potential; solid line) for $d\ll1$. Outside the forcing regions $\mathcal{R}$ and $\mathcal{L}$ (green dashed line), the leading-order contribution to the propagator $p(x,t|x',0)=\sum_{\sigma\sigma'}  p(x,\sigma,t|x',\sigma',0)$ (orange dash-dotted line) for long times $t\gg1$ is given by the solution in free space. Inside the regions, the dominant modification is a constant multiplicative shift.}
\end{figure}

Here we derive the scaling forms for the force-force correlation of a symmetric tracer given by Eq.~(22) of the main text, together with the scaling function $G$, as well as for the friction coefficients given by Eqs.~(23-26) of the main text. To do so, we use a systematic perturbative analysis for a piecewise linear tracer (see Fig.~\ref{fig:pert}). We start with a perturbation theory in $d/\ell_\text{p}$ and then consider the small $f_0$ (and $d$) limit. Once more, we make the equations dimensionless by rescaling $t\rightarrow\alpha^{-1}t$,
$x\rightarrow \ell_{\text{p}}x$ and $V\rightarrow T_{\text{eff}}V$. 

We begin with the $d\ll \ell_\text{p}$ limit and now turn to derive Eq.~(22) and its accompanying (correct) scaling function
\begin{align}
    G(y)&=\frac{1-(\frac{2\mu f_0 y}{v})^2}{[1-(\frac{\mu f_0}{v})^2]^2}~,\label{eq:scalingf}
\end{align}
which, upon rescaling the equations, becomes\begin{align}
    G(y)&=\frac{1-(2 f_0 y)^2}{(1-f_0^2)^2}~.\label{eq:scalingf1}
\end{align}
For any point $x$ outside the tracer sides, the leading-order contribution in $d$ to the propagator is given by the free propagator Eq.~\eqref{eq:fs}. However, inside the tracer sides, there is a correction due to the jump of the propagator at the side edges (see Fig.~\ref{fig:pert}). To obtain this correction, we integrate Eq.~\eqref{eq:rtpt} over a small region $[x-\varepsilon,x+\varepsilon]$, for any value of $x$, and take $\varepsilon\rightarrow0$. This results in
\begin{align}
\left.\left[1+\sigma f(x)\right]p(x,\sigma,s|x',\sigma')\right|_{-}^{+} & =\sigma\delta_{\sigma\sigma'}1_{xx'}~,\label{eq:interface1}
\end{align}
where $\left.g\right|_{-}^{+}\equiv g(x^{+})-g(x^{-})$
and $1_{xx'}=1$ if $x=x'$ and is $0$ otherwise. (Note that we have reinstated the explicit dependence on the initial values $x'$ and $\sigma'$ in $p$.) This means that $p(x,\sigma,s|x',\sigma')$ changes with $x$ in two ways:  It jumps discontinuously at the boundaries of the side regions---where $f$ is discontinuous---and at $x=x'$, according to Eq.~\eqref{eq:interface1}. It varies continuously elsewhere. Furthermore, in the $d\ll1$ limit, the continuous change within each of the narrow tracer sides $\mathcal{R}=[-d/2,d/2]$ and $\mathcal{L}=[L_\text{T}-d/2,L_\text{T}+d/2]$ can be neglected to leading order. This means that $p(x,\sigma,s|x',\sigma')$ is piece-wise constant within $\mathcal{R}$ and $\mathcal{L}$, with the plateau values being determined by the jumps at $x=x'$ and on the edges of those regions, using Eq.~\eqref{eq:interface1} and that the solution outside the tracer sides is $p^0(x,\sigma,s|x',\sigma')$. In sum, the leading-order expansion is
\begin{align}
    p(x,\sigma,s|x',\sigma')=\frac{p^0_\sigma(x,\sigma,s|x',\sigma')}{1+\sigma f(x)}+\mathcal{O}(d)~.\label{eq:pd}
\end{align}
Likewise, expanding Eq.~(11) gives
\begin{align}
    \rho_{\text{s}}(x',\sigma') = \frac{\rho_0}{2\left[1+\sigma'f(x')\right]} + \mathcal{O}(d)~.\label{eq:rd}
\end{align}
Eqs.~\eqref{eq:pd} and~\eqref{eq:rd} are then used to compute $C_{\mathcal{F}}(s)\equiv \mathcal{L}\left[C_{\mathcal{F}}(t)\right](s)$
and $\gamma(s)=\mathcal{L}\left[\gamma(t)\right](s)$, where $\mathcal{L}\left[g(t)\right](s)\equiv\int_{0}^{\infty}dte^{-st}g(t)$
is the Laplace transform. By definition, we get 
\begin{align}
C_{\mathcal{F}}(s) & =\sum_{\sigma\sigma'}\int dxdx'f(x)p(x,\sigma,s|x',\sigma')f(x')\rho_{\text{s}}(x',\sigma')~,\label{eq:FFin}\\
\gamma(s) & = \sum_{\sigma\sigma'}\int dxdx'f(x)p(x,\sigma,s|x',\sigma')\partial_{x'}\rho_{\text{s}}(x',\sigma')~.\label{eq:gamma}
\end{align}
Using the tracer symmetry, Eq.~\eqref{eq:FFin} becomes
\begin{align}
C_{\mathcal{F}}(s) & = 2\frac{\rho_0(f_0d)^2}{1-f_0^2}\sum_{\sigma\sigma'}\frac{1+\sigma'f_0}{2}p(x,\sigma,s|0,\sigma')|^{0}_{L_\text{T}}+\mathcal{O}(d^3)~,
\end{align}
Using Eqs.~\eqref{eq:fs},~\eqref{eq:pd} and~\eqref{eq:rd} and expanding for $s\ll1$ then gives
\begin{align}
    C_{\mathcal{F}}(s) & = \frac{\rho_0(f_0d)^2}{(1-f_0^2)^2} \left[L_\text{T}-\frac{1}{2}(L_\text{T}^2-4f_0^2)s^{1/2}\right]+\mathcal{O}(d^3,s^{3/2})~.\label{eq:Cfs}
\end{align}
In real time, the coefficients of $s^{1/2}$ in the $s\ll1$ expansions become the coefficients of $-(4\pi)^{-1/2}t^{-3/2}$ in the $t\gg1$ expansions, yielding
\begin{align}
    C_{\mathcal{F}}(t) & = \frac{\rho_{0}(f_0 d L_\text{T})^{2}}{4\pi^{1/2}t^{3/2}} G(L_\text{T}^{-1})+\mathcal{O}(d^3,t^{-5/2})~,\label{eq:Cfp}
\end{align}
where $G(y)$ is given by Eq.~\eqref{eq:scalingf1}. Upon restoring the dimensions, Eq.~\eqref{eq:Cfp} becomes Eq.~(22), with $G(y)$ given by Eq.~\eqref{eq:scalingf} as claimed. Additionally, Eq.~\eqref{eq:Cfs} provides the noise intensity
\begin{align}
    I&=\int_0^\infty dtC_\mathcal{F}(t)=C_{\mathcal{F}}(s=0) =\frac{\rho_0(f_0d)^2}{(1-f_0^2)^2}L_\text{T}+\mathcal{O}(d^3)~,
\end{align}
which remains positive irrespective of $L_\text{T}$.

We proceed to obtain the scaling forms for $\gamma_\text{p}(t)$, $\gamma_\text{a}(t)$ and $\gamma_\text{T}$ given in the main text. The change in the sign of $\gamma(s)$ can be observed by analysing the $f_0\ll1$ and $d\ll1$ limit. For this, we use the decomposition $\gamma(s)=\gamma_\text{p}(s)-\gamma_\text{a}(s)$ where $\gamma_\text{p}=\mathcal{L}[\gamma_\text{p}(t)](s)$ and $\gamma_\text{a}=\mathcal{L}[\gamma_\text{a}(t)](s)$. We also note that $p_\sigma=p_\sigma^0+\mathcal{O}(f_0)$ and $\rho_\text{s}(x',\sigma')=\rho_0/2+\mathcal{O}(f_0)$. Using Eqs.~(23) and~(24) in combination with Eq.~\eqref{eq:fs} then gives
\begin{align}
    \gamma_\text{p} &= \frac{C_{\mathcal{F}}(s)}{1-f_0^2} = \rho_0(f_0 d)^2\left(L_\text{T}-\frac{d}{3}\right)-\frac{1}{2}\rho_0(f_0 dL_\text{T})^2 s^{1/2}+\mathcal{O}(f_0^3,s^{3/2})~,\label{eq:gp1}\\
    \gamma_\text{a} &= 2\rho_0f_0^2d+\mathcal{O}(f_0^3,s^{3/2})~,\label{eq:ga1}
\end{align}
which leads to $\gamma_\text{a}(t)\sim\mathcal{O}(f_0^3)t^{-3/2}$ of the main text. Moreover, Eqs.~\eqref{eq:gp1}-\eqref{eq:ga1} yield
\begin{align}
    \gamma_\text{T}&=\int_0^\infty dt\gamma(t)=\gamma_\text{p}(s=0)-\gamma_\text{a}(s=0)=\rho_0(f_0 d)^2\left(L_\text{T}-\frac{d^2+6}{3d}\right)+\mathcal{O}(f_0^3)~,\label{eq:gT}
\end{align}
which, upon restoring dimensions, provides Eq.~(26). As seen in Eq.~\eqref{eq:gT}, $\gamma_\text{T}$ becomes negative for $L_\text{T}\lesssim(d^2+6)/3d$. Note that, since each term in Eq.~\eqref{eq:gT} is proportional to at least one power of $d$, the expansion is valid for $d\ll1$. In fact, its range of validity can be extended up to $d= \mathcal O(1)$.

\section{Finite temperature effects}

Here we discuss the case in which a finite temperature $T$ is retained in the generalized Langevin equation (1).

First, note that our results hold quantitatively in the limit in which $T$ is negligible for the RTPs, \emph{i.e.} $T\ll T_{\rm eff}$.
For asymmetric tracers, the thermal noise contribution is always negligible in the long-time limit, and the anomalous properties remain unchanged. For symmetric tracers, the diffusivity is simply shifted as: $D=T/\gamma_0+I/(\gamma_0+\gamma_T)^2$. 

Outside the small-$T$ limit, we expect the long-time tails to exhibit the same universal exponents.

Another interesting limit is when the coupling between the tracer and the particles is weak and smoothly varying, \emph{i.e.} $\mu |f(x)|\ll v$ and $|\partial_x f(x)|\ll\alpha/\mu$. Equation~(10) then yields the effective equilibrium distribution $\rho_\text{s}(x,\sigma)\simeq\rho_0 \exp[-\beta_\text{eff} V(x)$ $]/2$. Then, $F\simeq 0$, $\gamma(t)\simeq \beta_\text{eff}C_\mathcal{F}(t)$. The tracer is effectively coupled with two equilibrium baths of temperatures $T$ and $T_\text{eff}$. This leads to normal diffusion irrespective of the tracer shape, with an effective temperature for the tracer given by
\begin{align}
T_\text{T} & = \frac{D}{\gamma_t+\gamma_T}= \frac{\gamma_\text{T}}{\gamma_0+\gamma_\text{T}}T_\text{eff}+\frac{\gamma_0}{\gamma_0+\gamma_\text{T}}T~.
\end{align}
At sub-leading order in $f$ and $\partial_x f$, the distribution departs from its equilibrium approximation, and the anomalous properties are recovered.

\end{document}